\documentclass[12pt, draftclsnofoot, onecolumn]{IEEEtran} 

\usepackage{cite}

\ifCLASSINFOpdf
  \usepackage[pdftex]{graphicx}
\else
  \usepackage[dvips]{graphicx}
\fi

\usepackage[cmex10]{amsmath}
\usepackage{amssymb}

\usepackage{bm}
\usepackage{extarrows}

\usepackage{algorithm}
\usepackage{algorithmic}
\usepackage{xcolor}

\newcommand{\reffig}[1]{Fig. \ref{#1}}
\newcommand{\refeqn}[1]{(\ref{#1})}

\makeatletter

\newcommand{\Rmnum}[1]{\expandafter\@slowromancap\romannumeral #1@}
\makeatother

\newtheorem{theorem}{Theorem}

\newtheorem{lemma}{Lemma}

\newtheorem{proposition}{Proposition}
\newtheorem{corollary}{Corollary}
\newtheorem{definition}{Definition}

\usepackage{footnote}
\makesavenoteenv{tabular}

\begin{document}
%
\title{Widely-Linear Precoding for  Large-Scale MIMO with IQI: Algorithms and Performance Analysis}

 \author { \IEEEauthorblockN{ Wence Zhang, Rodrigo C. de Lamare, \IEEEmembership{Senior Member, IEEE}, Cunhua Pan, \\ Ming Chen, Jianxin Dai, Bingyang Wu and Xu Bao }


 \thanks{W. Zhang is with Jiangsu University, China. He was with CETUC, PUC-Rio, Brazil. (e-mail:wencezhang@ujs.edu.cn)}

\thanks{
R. C. de Lamare is with the University of York, UK, and PUC-Rio, Brazil. (e-mail:delamare@cetuc.puc-rio.br)}
\thanks{C.Pan is with School of Electronic Engineering and Computer Science, Queen Mary University of London, London E1 4NS, U.K. (e-mail:c.pan@qmul.ac.uk)}
\thanks{M. Chen and B. Wu are with National Mobile Communications Research Lab. (NCRL), Southeast University, China. (e-mail: \{chenming,wubingyang\}@seu.edu.cn).}

\thanks{J. Dai is with School of Science, Nanjing University of Posts and Telecommunications, China.           (email:daijx@njupt.edu.cn).}

\thanks{X. Bao is with Jiangsu University, China. (email:xbao@ujs.edu.cn).}

\thanks{Part of this work was published in Eusipco' 2014 and  ICC' 2015.}

}

\maketitle

\vspace*{-4em}
\begin{abstract}
\vspace*{-1em}
In this paper we study widely-linear precoding techniques to mitigate in-phase/quadrature-phase (IQ) imbalance (IQI) in the downlink of  large-scale  multiple-input multiple-output (MIMO) systems. We adopt a real-valued signal model which takes into account the IQI at the transmitter and then develop widely-linear zero-forcing (WL-ZF), widely-linear matched filter (WL-MF), widely-linear minimum mean-squared error (WL-MMSE) and widely-linear block-diagonalization (WL-BD) type precoding algorithms for both {\color{red} single- and multiple-antenna users.} We also present a performance analysis of WL-ZF and WL-BD. It is proved that without IQI, WL-ZF has exactly the same multiplexing gain and power offset as ZF, while when IQI exists,  WL-ZF achieves the same multiplexing gain as ZF with ideal IQ branches, but with a minor power loss which is related to the system scale and the IQ parameters. We also compare the performance of WL-BD with BD. The analysis shows that with ideal IQ branches, WL-BD has the same data rate as BD, while when IQI exists, WL-BD achieves the same multiplexing gain as BD without IQ imbalance. Numerical results verify the analysis and show that the proposed widely-linear type precoding methods significantly outperform their conventional counterparts with IQI and approach those with ideal IQ branches.
\end{abstract}

\vspace*{-1em}
\begin{IEEEkeywords}
\vspace*{-1em}
IQ imbalance, large-scale MIMO, widely-linear signal processing, downlink precoding
\vspace*{1em}
\end{IEEEkeywords}

%
\IEEEpeerreviewmaketitle

\vspace*{-2em}
\section{Introduction}

Wireless communications systems have undergone tremendous development during the past decades. In order to meet the increasing demands for data services, many new techniques have been proposed, among which {\color{red} multiple-input multiple-output} (MIMO) techniques play an important role. In the 5th generation (5G) of mobile communication systems, one of the key techniques will be large-scale MIMO, which employs a large number of antennas at the base station (BS) with centralized or distributed antenna systems to provide extremely high data rates with improved quality of service (QoS)\cite{Rusek2013MSP,Larsson2014MCOM}.

{ One of the main performance constraints of large-scale MIMO systems comes from the impairments resulting from hardware\cite{Lamare2013URSI,Zhang2015TCOM}. Since large-scale MIMO systems employ a large number of antennas, cheaper hardware is preferable in order to reduce the cost, which may cause severe hardware imperfection, e.g., the in-phase and quadrature-phase (IQ) imbalance (IQI)\cite{Hakkarainen2013JCN}. 
Modern transceivers usually use the direct conversion structure which contains two branches to process the real and imaginary components of the baseband signals, i.e., the in-phase (I) branch and the quadrature-phase (Q) branch. The IQI exists when there is a gain difference between the two branches and/or the phase difference is not exactly $90^\circ$. The IQI can be present at both the transmitter and the receiver, according to many studies\cite{Tarighat2007TWC,Zhang2014LCOMM,Zhang2015ICC2}.}

{\color{red} One way of handling IQI is to estimate the IQ parameters and compensate for them (see \cite{Darsena2014EPC,Zhang2015ICC2,Darsena2015ICC} and references therein)}. However, the IQ parameters are usually mingled with the channel coefficients, and thus are difficult to obtain, especially for large-scale MIMO systems, where the number of IQ parameters is proportional to the system size and thus the estimation and compensation for IQI can be very computationally expensive.

{ {\color{red} Widely-linear approaches have long been used for non-circular signal processing in MIMO systems \cite{Sterle2007TSP,Darsena2012ISCCSP,Darsena2013ISWCS} and  have been recently adopted to deal with IQI \cite{Hakkarainen2013CROWNC, Hakkarainen2013RWS,Hakkarainen2013JCN,Hakkarainen2014GCW, Zhang2014EUSIPCO, Zarei2014GCW,Zhang2015ICC,Kolomvakis2015ICC, Zarei2016TWC, Hakkarainen2016TWC}}.  In the {\it uplink}, the impact of IQI on the response pattern of large antenna arrays is studied in \cite{Hakkarainen2013RWS}. The work in \cite{Hakkarainen2014GCW} describes an equivalent interference model to study the impact on {\color{red} orthogonal frequency division multiple access (OFDMA)} large-scale MIMO systems and devised a receiver based on widely-linear signal processing, which is extended in \cite{Hakkarainen2016TWC} to scenarios with external interference. In \cite{Kolomvakis2015ICC}, the authors investigated the impact of IQI on the performance of uplink Massive MIMO systems with maximum-ratio combining (MRC) receivers, and showed that IQI can substantially degrade the performance of MRC receivers. The study in \cite{Kolomvakis2015ICC} also proposed a low-complexity IQI compensation scheme. In order to suppress the impact of IQI, a data-aided widely-linear minimum mean square error (MMSE) receiver is proposed in \cite{Hakkarainen2013ISWCS}, and  an IQI aware receiver was designed in \cite{Hakkarainen2013JCN} for the large-scale MIMO uplink based on the minimum variance distortionless response (MVDR) criterion. In \cite{Zarei2016TWC}, a widely-linear MMSE receiver is proposed, the performance of which is shown to be close to the linear MMSE receiver in an ideal system without IQI. {\color{red} Although duality exists in the uplink and downlink\cite{Jindal2004TIT}, these results are different from that in the downlink, because of hardware mismatch which results in different channel coefficients in the uplink and the downlink\cite{Zhang2015TCOM},  and a power constraint imposed on the downlink precoder design which does not exist in the uplink detection. }

To the best of the authors' knowledge, there are only a few related works in the {\it downlink} \cite{Hakkarainen2013CROWNC,Zhang2014EUSIPCO,Zarei2014GCW,Zhang2015ICC}. The study in \cite{Hakkarainen2013CROWNC} uses an augmented representation to maximize the power of the desired transmit signal when IQI presents.  Reduced-rank widely-linear precoders were devised in \cite{Zhang2014EUSIPCO} for single-antenna users to alleviate the impact of IQI as well as to reduce the computational complexity. A similar work was reported later in \cite{Zarei2014GCW}, which studied the impact of IQI and proposed a widely-linear regularized zero-forcing (RZF) precoding scheme.  In \cite{Zhang2015ICC}, we extended our previous work in \cite{Zhang2014EUSIPCO} to scenarios with multiple-antenna users in the large-scale MIMO downlink and developed novel widely-linear block diagonal (BD) type precoders.

In this paper, our previous work \cite{Zhang2014EUSIPCO,Zhang2015ICC} is extended to give a comprehensive study on widely-linear precoding algorithms for the large-scale MIMO downlink with IQI for users with both single and multiple antennas. For large-scale MIMO systems with single-antenna users, non-linear precoding schemes, e.g., vector perturbation (VP) precoding \cite{Hochwald2005TCOM} and Tomlinson-Harashima precoding \cite{Tomlinson1971EL,Harashima1972TCOM}, usually have better performance than linear precoding schemes. However, linear precoding schemes, such as matched filter (MF) (also referred to as maximum ratio transmission \cite{Lo1999TCOM}), zero-forcing (ZF) and MMSE \cite{Peel2005TCOM}, have much lower complexity compared with the nonlinear precoding schemes and thus draw great research interest\cite{Rusek2013MSP,Larsson2014MCOM}. Generally, ZF and MMSE perform better than MF, but with a comparatively higher computational cost due to the matrix inversion involved.
}

Most current studies on large-scale MIMO downlink have considered single-antenna users. However, it is well known that with more receive antennas at the user equipment (UE), the quality of service  (QoS) of each user can be significantly improved. In fact, the long term evolution (LTE) and LTE-Advanced (LTE-A) standards can support UEs with multiple antennas\cite{Liu2012MCOM}.  In terms of downlink precoding schemes, block-diagonalization (BD) type precoding has been widely considered for parallel transmission of multiple data streams for users with multiple antennas\cite{Spencer2004TSP,Choi2004TWC,Stankovic2008TWC,Sung2009TCOMM,Zu2013TCOMM}.
In \cite{Stankovic2008TWC}, the regularized BD (RBD) precoding has been proposed which outperforms conventional BD in \cite{Spencer2004TSP,Choi2004TWC}, by taking into account both the inter-user interference and noise. { To reduce the complexity, the work in \cite{Sung2009TCOMM} has devised the generalized MMSE channel inversion (GMI) by replacing the  singular value decomposition (SVD) operation in BD and RBD with a matrix inversion and QR decomposition. This scheme has been further modified to obtain the simplified GMI (S-GMI) technique in \cite{Zu2013TCOMM}}.

{
In contrast to \cite{Hakkarainen2013RWS,Hakkarainen2014GCW,Hakkarainen2016TWC,Kolomvakis2015ICC,Hakkarainen2013ISWCS,Hakkarainen2013JCN,Zarei2016TWC}, in this work we study the design and performance analysis for {\it downlink} precoding in large-scale MIMO systems with transmitter IQI, for both single- and {\color{red} multiple-antenna users}. We first adopt a useful mapping function reported in \cite{Telatar1999ETT,Hellings2015TSP}, which transforms complex-valued vectors and matrices into real-valued expressions and  helps to set up an equivalent real-valued signal model with consideration of IQI. Based on this real-valued signal model, we develop widely-linear ZF (WL-ZF) and widely-linear MF (WL-MF) and widely-linear MMSE (WL-MMSE) precoders, which are suited for single-antenna terminals. Unlike \cite{Hakkarainen2013CROWNC,Zhang2014EUSIPCO,Zarei2014GCW} where only single-antenna users are considered, we propose widely-linear BD (WL-BD) type precoding algorithms, i.e., WL-BD, widely-linear RBD (WL-RBD) and widely-linear S-GMI (WL-S-GMI),  for users equipped with multiple antennas. A performance analysis is carried out for the ZF and BD versions of these proposed precoding schemes, which captures the essential advantages of widely-linear precoding approaches. By utilizing an affine approximation of the sum data rate developed in \cite{Shamai2001TIT}, the mathematical expression for the sum data rate of WL-ZF is derived. Moreover, we also compare ZF and WL-ZF in terms of both multiplexing gain and power offset, where the IQ parameters are treated as random variables in the analysis, which is different from \cite{Zarei2014GCW,Hakkarainen2016TWC,Zarei2016TWC}, where the IQ parameters are fixed. For WL-BD, the sum data rate and multiplexing gain are derived and compared with those of BD.   We also give simulation results to show the impact of IQI in large-scale MIMO systems, as well as the performance of the proposed widely-linear precoding schemes.


The main contributions of this paper are summarized as follows:
\begin{itemize}
  \item {\color{red} We extend previous work of WL-ZF, WL-MF and WL-MMSE precoding for large-scale MIMO systems with IQI and single-antenna users in \cite{Zhang2014EUSIPCO} to cases with multiple-antenna users and propose WL-BD, WL-RBD and WL-S-GMI algorithms based on an equivalent real-valued signal model.}
  \item For WL-ZF, we show that it has the same multiplexing gain as that of ZF with ideal IQ branches. The achieved multiplexing gain equals the number of users. Compared with ZF without IQI, WL-ZF has a power offset loss around $\log_2 [1+4\sigma_g^2\beta]$, where $\beta$ is the ratio of the number of users to that of the transmit antennas and $\sigma_g^2$ is the variance of the gain difference between I and Q branches.
  \item For WL-BD, we prove that when there is no IQI, it achieves the same sum data rate as BD, while the WL-BD precoder in the presence of IQI has the same multiplexing gain as that of BD with ideal IQ branches. Moreover, compared with BD the  increased complexity of WL-BD is very small.
\end{itemize}
}


This paper is organized as follows. The system model is described in Section II. In Section III, the proposed widely-linear precoding algorithms are introduced. In Section IV, the performance analysis of WL-ZF and WL-BD is carried out. The numerical results are given in Section V and conclusions are drawn in Section VI.

\emph{Notation:} $\mathbb C^{N}$ and $\mathbb C^{N\times M}$ denote the sets of $N$-dimensional complex vectors and  $N \times M$ complex matrices, respectively; $\mathbb R^{N}$ and $\mathbb R^{N\times M}$ denote the sets of $N$-dimensional real vectors and  $N \times M$ real matrices, respectively; $(\cdot)^*$ is the complex conjugate; $\bm{I}_N$ denotes an $N \times N$ identity matrix; ${\cal{CN}} (\bm{\theta},\bm{\Sigma})$ denotes circularly symmetric complex Gaussian distribution with mean $\bm{\theta}$ and covariance $\bm{\Sigma}$; $U(a, b)$ denotes the uniform distribution and $\mathcal N (a, b)$  denotes the Gaussian distribution, where $a$ and $b$ are the  mean and variance, respectively; $\text{diag}\{a_1,\cdots, a_K\}$ denotes a diagonal matrix with diagonal entries given by $a_1,\cdots, a_K$; { \color{red} $\mathbb E\{\cdot\}$ denotes the mathematical expectation and $\text{Tr}[\bm A]$ denotes the trace of a matrix $\bm A$.}

\vspace*{-1em}
\section{System Model}

Consider the downlink of a large-scale MIMO system with one BS and  $K$ users. The BS is equipped with $N$ antennas, and there are $M_k$ antennas at the $k$-th user. Define $M=\sum_{k=1}^K M_k$ as the total number of antennas at all the users. {Consider the IQI at all the $N$ transmit antennas at BS\footnote{\color{red}  The IQI at the user's receiver only degrades its own signal and can be addressed individually by IQI compensation techniques\cite{Zhang2015ICC2}. In contrast, the IQI at the BS affects all the users and is severe in large-scale MIMO systems due to the potential use of cheap hardware for cost issues. Therefore, we only consider the IQI at the BS in this paper.}, and for the $n$-th antenna, the transmit symbol $x_n$ is corrupted by IQI as $a_{n1} x_n + a_{n2} x_n^*$, where $a_{n1}$ and $a_{n2}$ are the IQ parameters of the $n$-th antenna that are modeled as \cite{Tarighat2007TWC}:
\begin{equation} \label{eqn:a1_a2}
  \begin{split}
    a_{n1} =\frac{1}{\sqrt{1+\sigma_g^2}} [\cos(\theta_n / 2) + {\text j} g_n \sin(\theta_n/2)], \quad
    a_{n2} = \frac{1}{\sqrt{1+\sigma_g^2}} [g_n \cos(\theta_n / 2) - {\text j} \sin(\theta_n/2)],
  \end{split}
\end{equation}
where $\theta_n \sim U(0, \sigma_{\theta}^2)$  and $g_n \sim \mathcal{N}(0,\sigma_g^2)$ are the relative phase and gain mismatches between the IQ branches of the $n$-th transmit antenna, respectively. The IQ parameters are normalized so that they do not change the average signal power. {\color{red}  A proof of the selection of the normalization factor is given in Appendix \ref{app:norm}}. In \refeqn{eqn:a1_a2}, $\theta_n = 0^\circ$ and $g_n = 0$ represent the ideal case with no IQI.
Note that  although the Gaussian and uniform distributions are considered for modeling IQ parameters, the proposed algorithms and the performance analysis are valid for other distributions.}

{
If we consider the IQI at BS,  the received signal $\bm y_k \in \mathbb C^{M_k}$ at the $k$-th user is given by
\begin{equation}\label{eqn:rx_sig_k}
  \bm y_k = \bm H _k \bm A_1 \bm  x + \bm H_k \bm A_2 \bm x^* + \bm n_k,
\end{equation}
where $\bm H_k \in \mathbb C^{M_k \times N}$ is the downlink channel matrix of the $k$-th user, the elements of which are independent and identically distributed (i.i.d.) {\color{red} Gaussian random variables} with zero-mean and unit variance; $\bm A_1 = {\text {diag}}\{a_{11},\cdots,a_{N1}\}$ and $\bm A_2 = {\text {diag}}\{a_{12},\cdots,a_{N2}\}$; $\bm n_k \sim \mathcal{CN} (\bm 0,\sigma_n^2 \bm I_{M_k})$ is the noise vector at the receiver; $\bm x \in \mathbb C^{N}$ is the transmit signal vector after precoding. Here we consider a narrow-band single-carrier system for simplicity and the extension to multi-carrier systems remains open for future work.}

Let $L_k$ be the number of data streams of user $k$ and
$\bm P_k \in \mathbb C^{N\times L_k}$, $\bm s_k \in \mathbb C^{L_k}$ be the precoder and the transmit signal vector for the $k$-th user, respectively. Denote
$\bm P = [\bm P_1, \cdots, \bm P_K]$, $\bm s = [\bm s_1^{\text T}, \cdots, \bm s_K^{\text T}]^{\text T}$, and we have $\bm x = \bm P \bm s$.
 {  \color{red} Note that in contrast to single-antenna users using ZF, MMSE or MF, in the case of multiple-antenna users using BD type precoders, a receive filter matrix is generally required to decode the multiple streams which is designed together with the precoder, as will be detailed later in Section III.}

 { In this paper, we assume that the transmitter has perfect channel state information and the transmit signals for different users are i.i.d {\color{red} Gaussian random variables} with zero-mean and unit variance}, i.e.,
 $\forall k\neq j$, $\mathbb E \{\bm s_k \bm s_j^{\text H}\} = \bm 0$, and $\mathbb E \{\bm s_k \bm s_k^{\text H}\} = \bm I_{L_k}$. We also assume there is a transmit power constraint, i.e.,
 \begin{equation}\label{eqn:pow_cons}
   \mathbb E \{\| \bm P \bm s\|^2\} = P_{\text T}.
 \end{equation}

It can be seen from \refeqn{eqn:rx_sig_k} that the transmit signal vector is corrupted by its complex conjugate. In the frequency domain, a mirror frequency component is introduced due to the IQI. One possible way of handling such IQI resorts to estimation of the corresponding IQ parameters and pre-compensation for the IQI \cite{Tarighat2007TWC,Zhang2014LCOMM}. Since the signal model of \refeqn{eqn:rx_sig_k} gives rise to non-circular data which can be exploited by widely-linear processing, IQI can also be tackled by widely-linear approaches \cite{Hakkarainen2013JCN,Zhang2014EUSIPCO}.

In what follows, we will devise and carry out a performance analysis of widely-linear precoding schemes, which are able to mitigate the IQI without significantly increasing the computational complexity.


\vspace*{-1em}
\section{Proposed widely-linear Precoding Algorithms}

{ In this section, we employ a useful transformation, i.e., the $\mathcal T$-transform from \cite{Telatar1999ETT,Hellings2015TSP}, which represents complex-valued matrices and vectors using their real-valued equivalents. Then we employ the $\mathcal T$-transform to develop an equivalent real-valued signal model, which helps to design widely-linear precoding schemes.}  Several widely-linear precoding algorithms such as WL-ZF, WL-MMSE, WL-BD, WL-RBD and WL-S-GMI are then developed.

\vspace*{-1em}
\subsection{Real-Valued Signal Model}
{
A mapping function of ${\mathbb{C}^n} \to {\mathbb{R}^{2n}}$   and ${\mathbb{C}^{m \times p}} \to {\mathbb{R}^{2m \times 2p}}$, namely the $\mathcal T$-transform,  is defined as:
\begin{equation}\label{eqn:transform}
\mathcal T({\bm{x}}) =  \begin{bmatrix}
  \operatorname{Re} ({\bm{x}})  \\
  \operatorname{Im} ({\bm{x}})
\end{bmatrix} , \mathcal T({\bm{X}}) = {\begin{bmatrix}
  {\operatorname{Re} ({\bm{X}})}&{ - \operatorname{Im} ({\bm{X}})} \\
  {\operatorname{Im} ({\bm{X}})}&{\operatorname{Re} ({\bm{X}})}
\end{bmatrix}} ,
\end{equation}
where $\operatorname{Re} ( \cdot )$  and $\operatorname{Im} ( \cdot )$  represent the real and imaginary parts of a vector or a matrix, respectively. The $\mathcal T$-transform sets up a relationship between the complex-valued matrices and their real-valued counterparts. It is very useful for design and performance analysis of widely-linear precoders. Some properties of the $\mathcal T$-transform are summarized in Lemma \ref{lem:t_trans}, Corollary \ref{lem:t_trans_2} and \ref{lem:t_unitary}. More information on this transform can be found in \cite{Telatar1999ETT,Hellings2015TSP}.

\begin{lemma} [Lemma 1,\cite{Telatar1999ETT}] \label{lem:t_trans}
The following equations hold if the corresponding matrix or vector operation is valid:
\begin{equation}\label{eqn:t_prop}
\begin{matrix}
  &{\mathcal T({\bm{AB}}) = \mathcal T({\bm{A}})\mathcal T({\bm{B}})}, {\mathcal T({{\bm{A}}^{ - 1}}) = {{[\mathcal T({\bm{A}})]}^{ - 1}}},
  {\mathcal T({\bm{A}} + {\bm{B}}) = \mathcal T({\bm{A}}) + \mathcal T({\bm{B}})}, \\ &{\mathcal T({{\bm{A}}^{\rm{H}}}) = {{[\mathcal T({\bm{A}})]}^{\rm{H}}}},
  {\mathcal T({\bm{x}} + {\bm{y}}) = \mathcal T({\bm{x}}) + \mathcal T({\bm{y}})}, {\mathcal T({\bm{Ax}}) = \mathcal T({\bm{A}})\mathcal T({\bm{x}})}.
\end{matrix}
\end{equation}
\end{lemma}
\begin{IEEEproof}
Please refer to \cite{Telatar1999ETT}.
\end{IEEEproof}

\begin{corollary}\label{lem:t_trans_2}
Denote $\bm E_N$ and $\bar{\bm I}_N$ as
\begin{equation*}
  \bm E_N = \begin{bmatrix} \bm I_N & \\ & -\bm I_N  \end{bmatrix}, \bar{\bm I}_N = \begin{bmatrix} & \bm I_N \\ \bm I_N &  \end{bmatrix}.
\end{equation*}
Then we have
\begin{equation}\label{eqn:en_in}
\begin{split}
  \bm E_N \mathcal T(\bm H) \bm E_N  = \bar{\bm I}_N \mathcal T(\bm H) \bar{\bm I}_N = \mathcal T(\bm H^*), \quad
  \bar{\bm I}_N \mathcal T(\bm H) \bm E_N  = \begin{bmatrix} \text{Im}(\bm H) & -\text{Re}(\bm H) \\ \text{Re}(\bm H) & \text{Im}(\bm H) \end{bmatrix}.
  \end{split}
\end{equation}
If $\bm H$ and $\bm G$ are $2N \times 2N$ Hermitian matrices, then we have
\begin{equation}\label{eqn:t_her}
    \text{Tr}[ \mathcal T(\bm H)] = 2\text{Tr}[ \bm H],     \text{Tr}[\mathcal T(\bm H)\mathcal T(\bm G)] = 2\text{Tr}[ \bm H \bm G], \quad
    \text{Tr}[ \mathcal T(\bm H) \bm E_N ] = \text{Tr}[ \mathcal T(\bm H) \bar{\bm I}_N ] = 0.
\end{equation}
\end{corollary}
\begin{IEEEproof}
Equation \refeqn{eqn:en_in} is proved using results in Lemma \ref{lem:t_trans}, while \refeqn{eqn:t_her} follows the fact that the diagonal elements of a Hermitian matrix are real-valued and the trace of the product of two Hermitian matrices is also real-valued.
\end{IEEEproof}

\begin{corollary}\label{lem:t_unitary}
The $\mathcal T$-transform of a complex-valued unitary matrix is a real-valued orthogonal matrix. Moreover, let $\bm X_{\text r}$ be a permutation of rows of a matrix $\bm X$, and the SVD of $\bm X$ and $\bm X_{\text r}$ are given by $\bm X = \bm U \bm \Sigma \bm V^{\text H}$, $\bm X_{\text r} = \bm U_{\text r} \bm \Sigma_{\text r} \bm V_{\text r}^{\text H}$, respectively. Then we have that $\bm U_{\text r}$ is a permutation of rows of $\bm U$, $\bm \Sigma_{\text r} = \bm \Sigma$ and $\bm V_{\text r} = \bm V$.
\end{corollary}
\begin{IEEEproof}
It is straightforward to achieve this corollary from Lemma \ref{lem:t_trans}. More detailed discussion is provided in \cite{Hellings2015TSP}.
\end{IEEEproof}

By applying the $\mathcal T$-transform to \refeqn{eqn:rx_sig_k}, the following real-valued signal model is achieved for the $k$-th user:
\begin{equation}\label{eqn:rx_sig_k_ex}
\begin{split}
 \tilde {\bm y}_k = \mathcal T(\bm y_k)
                 = \mathcal T(\bm H _k)[ \mathcal T(\bm A_1) + \mathcal T(\bm A_2)\bm E_N ] \mathcal T(\bm  x) + \mathcal T( \bm n_k)
                  \triangleq \tilde{\bm H}_k \tilde {\bm A}\tilde {\bm x} + \tilde {\bm n}_k,
  \end{split}
\end{equation}
where $\tilde{\bm H}_k = \mathcal T(\bm H_k)$, $\tilde{\bm A} = \mathcal T(\bm A_1) + \mathcal T(\bm A_2)\bm E_N$, $\tilde{\bm x} = \mathcal T(\bm x)$ and $\tilde{\bm n}_k = \mathcal T(\bm n_k)$.
Denoting $\tilde {\bm y} = [\tilde{\bm y}_1^{\text T},\cdots, \tilde{\bm y}_K^{\text T}]^{\text T}$, $\tilde{\bm H} = [\tilde{\bm H}_1^{\text T} ,\cdots, \tilde{\bm H}_K^{\text T}]^{\text T}$, $\tilde {\bm n} = [\tilde{\bm n}_1^{\text T},\cdots, \tilde{\bm n}_K^{\text T}]^{\text T}$, $\tilde{\bm P_i} = \mathcal T(\bm P_i)$ and $\tilde{\bm s}_i = \mathcal T(\bm s_i)$, we have
\begin{equation}\label{eqn:rx_sig_all}
 \tilde {\bm y} = \tilde{\bm H} \tilde {\bm A} \tilde{\bm x} + \tilde {\bm n} = \tilde{\bm H} \tilde {\bm A} \tilde{\bm P} \tilde{\bm s} + \tilde {\bm n},
\end{equation}
where $\tilde{\bm P} = [\tilde{\bm P}_1 ,\cdots, \tilde{\bm P}_K]$ and $\tilde{\bm s} = [\tilde{\bm s}_1^{\text T} ,\cdots, \tilde{\bm s}_K^{\text T}]^{\text T}$. In \refeqn{eqn:rx_sig_all}, $\mathbb E \{ \tilde{\bm s} \tilde{\bm s}^{\text T}\} = \frac{1}{2} \bm I$ and $\mathbb E \{ \tilde{\bm n} \tilde{\bm n}^{\text T}\} = \frac{1}{2}\sigma_n^2 \bm I$.}

The real-valued signal model in \refeqn{eqn:rx_sig_all} has a linear representation. Compared with the complex-valued signal model in \refeqn{eqn:rx_sig_k}, the system dimension is doubled, even though the operations with matrices and vectors are simplified due to the use of real-valued samples.
Based on \refeqn{eqn:rx_sig_all}, any precoding {\color{red} scheme} (e.g., MF, ZF, MMSE, BD, RBD and S-GMI, etc.) can be developed to cope with the IQI by treating $\tilde{\bm H} \tilde{\bm A}$ as the effective channel matrix. { \color{red} Since the real and imaginary parts of the transmit signals are processed separately, these schemes are referred to as ``widely-linear precoding'' schemes \cite{Darsena2012ISCCSP,Hakkarainen2013ISWCS,Zhang2014EUSIPCO,Zarei2014GCW}.}

In order to show how to design widely-linear precoding schemes, we derived several algorithms and focus our analysis on two typical examples: WL-ZF for single-antenna users and WL-BD for multiple-antenna users, the performance analysis of which are carried out in Section IV.

\vspace{-1em}

\subsection{Widely-Linear Precoding for Single-Antenna Users}

When each user is equipped with one antenna, we have $M_k = 1$ and $M=K$. This is the typical case as we studied in our previous work in \cite{Zhang2014EUSIPCO} or a similar work in \cite{Zarei2014GCW}. In this paper, we will focus on performance analysis in the next Section.

{
From \refeqn{eqn:rx_sig_all}, by treating the real and imaginary components as independent virtual users, the precoding matrix of  WL-ZF precoding is given by
\begin{equation}\label{eqn:wl_zf}
  {{\tilde{\bm P}}_{{\text{WL-ZF}}}} = \sqrt {{\lambda_{\text{WL-ZF}} }} {(\tilde{\bm H}\tilde {\bm A})^{\rm{T}}}{({\tilde{\bm H}}\tilde{\bm A} \tilde{\bm A}^{\text T}{{\tilde{\bm H}}^{\rm{T}}})^{ - 1}},
\end{equation}
where $\lambda_{\text{WL-ZF}}$ is the power normalization factor, which is obtained from \refeqn{eqn:pow_cons} and  given by
\begin{equation*}
  \lambda_{\text{WL-ZF}} = \frac{2P_{\text T}}{ \mathbb E\{ \text{Tr}[(\tilde{\bm H} \tilde{\bm A} \tilde{\bm A}^{\text T} \tilde{\bm H}^{\text{T}})^{-1}]\}},
\end{equation*}
in which the term $2P_{\text T}$ comes from $\mathbb E \{\tilde{\bm s} \tilde{\bm s}^{\text T}\} = \frac{1}{2}\bm I$. Note that in order to simplify the analysis the power normalization factor is calculated based on the expectation of $\text{Tr}[(\tilde{\bm H} \tilde{\bm A} \tilde{\bm A}^{\text T} \tilde{\bm H}^{\text{T}})^{-1}]$ as in \cite{Hoydis2013JSAC,Zhang2015TCOM}, other than the instantaneous value. However, when $N$ and $K$ is large the latter one will generally converge to its expected value almost surely \cite{Bai2010SA}. Therefore, the analysis in the following sections also gives close approximation when instantaneous channel information is considered.

Similarly, the precoding  matrices of WL-MF and WL-MMSE are obtained as \cite{Zhang2014EUSIPCO,Zarei2014GCW}:
\begin{equation}\label{eqn:wl_mf_mmse}
\begin{split}
  {{\tilde{\bm P}}_{{\text{WL-MF}}}} = \sqrt {{\lambda_{\text{WL-MF}} }} {(\tilde{\bm H}\tilde {\bm A})^{\rm{T}}},  \quad
  {{\tilde{\bm P}}_{{\text{WL-MMSE}}}} = \sqrt {{\lambda_{\text{WL-MMSE}} }} {(\tilde{\bm H}\tilde {\bm A})^{\rm{T}}}{({\tilde{\bm H}}\tilde{\bm A} \tilde{\bm A}^{\text T}{{\tilde{\bm H}}^{\rm{T}}} + \rho \bm I_{2K}  )^{ - 1}},
  \end{split}
\end{equation}
where $\rho = \frac{M\sigma_n^2}{P_{\text T}}$ \cite{Joham2005TSP} and
\begin{equation*}
\begin{split}
  \lambda_{\text{WL-MF}} = \frac{2P_{\text T}}{ \mathbb E \{\text{Tr}[ \tilde{\bm H} \tilde{\bm A} \tilde{\bm A}^{\text T} \tilde{\bm H}^{\text T} ]\}},
  \lambda_{\text{WL-MMSE}} = \frac{2P_{\text T}}{ \mathbb E \{\text{Tr}[ \tilde{\bm H} \tilde{\bm A} \tilde{\bm A}^{\text T}  \tilde{\bm H}^{\text T} (\tilde{\bm H} \tilde{\bm A} \tilde{\bm A}^{\text T} \tilde{\bm H}^{\text{T}} + \rho \bm I_{2K})^{-2}] \} }.
\end{split}
\end{equation*}
{\color{red} Note that although $2M$ streams are transmitted, $\rho$ has the same expression as for $M$ streams, because both the average transmit power and the noise power for each stream is halved.}
}

\subsection{Widely-Linear BD-Type Precoding for Multiple-Antenna Users}

In this subsection, we describe how to design BD-type precoding algorithms using the proposed real-valued signal model.

\subsubsection{WL-BD Precoding}
 The WL-BD precoding matrix of the $k$-th user is composed of four parts:
  $$
  \tilde{\bm P}_k = \sqrt{\lambda_{\text {WL-BD}}} \tilde{\bm P}_{k1} \tilde{\bm P}_{k2} \tilde{ \bm \Gamma}_k,
  $$
  where $\tilde{\bm P}_{k1} \in \mathbb C^{2N \times D_k}$,  $\tilde{\bm P}_{k2} \in \mathbb C^{D_k \times 2L_k}$ (the value of $D_k$ depends on how $\tilde{\bm P}_{k1}$ and $\tilde{\bm P}_{k2}$ are obtained) will be given in the following; $\tilde{\bm \Gamma}_{k} \in \mathbb R^{2L_k \times 2L_k}$ is the diagonal power loading matrix; $\lambda_{\text {WL-BD}}$ is the power factor to fulfill the transmit power constraint, i.e., $\mathbb E \{{\text {Tr}} [ \tilde {\bm P} \tilde{\bm s} \tilde {\bm s}^{\text T} \tilde{\bm P}^{\text T} ] \}= P_{\text T}$.

Let us exclude the $k$-th user's channel matrix and define $\tilde {\bm H}_{-k} \in \mathbb R^{2M_{-k}\times 2N}$  as
\begin{equation}\label{eqn:H_k_c}
  \tilde {\bm H}_{-k} = [\tilde{\bm H}_1^{\text T}, \cdots, \tilde{\bm H}_{k-1}^{\text T},\tilde{\bm H}_{k+1}^{\text T},\cdots, \tilde{\bm H}_K^{\text T}]^{\text T},
\end{equation}
where $M_{-k} = M-M_k$. {\color{red} Consequently, we have $\tilde {\bm H}_{-k}\tilde{\bm A} = [(\tilde{\bm H}_1\tilde{\bm A})^{\text T}, \cdots, (\tilde{\bm H}_{k-1}\tilde{\bm A})^{\text T}, (\tilde{\bm H}_{k+1}\tilde{\bm A})^{\text T}, $ $\cdots, (\tilde{\bm H}_K\tilde{\bm A})^{\text T}]^{\text T}$.
}

The precoding matrix $\tilde{\bm P}_{k1}$ is chosen to be in the null space of $\tilde{\bm H}_{-k}\tilde{\bm A}$, i.e., $\tilde{\bm H}_{-k} \tilde{\bm A} \tilde{\bm P}_{k1} = \bm 0$. Therefore, $\tilde{\bm P}_{k1}$  is chosen as the right singular vectors corresponding to the zero singular values of $\tilde{\bm H}_{-k}\tilde{\bm A}$ \cite{Spencer2004TSP,Choi2004TWC}.  Let  $\tilde{\bm H}_{-k}\tilde{\bm A} = \tilde{\bm U}_{-k} \tilde {\bm \Sigma}_{-k} \tilde{\bm V}_{-k}^{\text H} =\tilde{\bm U}_{-k} \tilde {\bm \Sigma}_{-k} [\tilde {\bm V}_{-k1}, \tilde{\bm V}_{-k0} ]^{\text H}$ be the SVD of $\tilde{\bm H}_{-k}\tilde{\bm A}$, where $\tilde{\bm V}_{-k0} \in \mathbb C^{2N\times D_k}$ contains the right singular vectors corresponding to the zero singular values of $\tilde{\bm H}_{-k}\tilde{\bm A}$. Then we have $\tilde{\bm P}_{k1}=\tilde{\bm V}_{-k0}$.

Consequently, the effective channel matrix for the $k$-th user is defined as:
\begin{equation} \label{eqn:h_k_e}
\tilde {\bm H}_{\text e _k} = \tilde {\bm H}_k \tilde{\bm A} \tilde{\bm P}_{k1}.
\end{equation}
The second component of the precoder can be obtained by applying SVD to the effective channel matrix as $\tilde {\bm H}_{\text e _k} = \tilde {\bm U}_{k} \tilde{\bm \Sigma}_k \tilde{\bm V}_k^{\text H}$. {\color{red} Then we have $\tilde{\bm P}_{k2} = \tilde{\bm V}_k$ and the corresponding receive filter matrix is  $\tilde{\bm G}_k = \tilde {\bm U}_{k}^{\text H}$}.

\subsubsection{WL-RBD Precoding}
Instead of totally eliminating the inter-user interference, $\tilde{\bm P}_{k1}$ can also be calculated according to the MMSE criterion\cite{Stankovic2008TWC}, which is given by
\begin{equation*}
  \tilde{\bm P}_{k1} = \tilde{\bm V}_{-k}  ( \tilde {\bm \Sigma}_{-k}^{\text H} \tilde {\bm \Sigma}_{-k} + \rho \bm I_{2N} )^{-1/2}.
\end{equation*}
Once $\tilde{\bm P}_{k1}$ is obtained, the effective channel matrix can be obtained as in \refeqn{eqn:h_k_e}. $\tilde{\bm P}_{k2}$ and $\tilde{\bm G}_k$ can be then obtained through SVD of $\tilde{\bm H}_{\text{e}_k}$.  This precoding method is referred to as WL-RBD in this paper.

\subsubsection{WL-S-GMI Precoding}
In fact,  when $N$ and $M$ are large it could be computationally expensive for BD and RBD  to calculate $K$ SVDs to get all $\tilde{\bm P}_{k1}$'s, $k=1,\ldots,K$.  Therefore, an alternative approach based on the GMI technique was proposed in \cite{Sung2009TCOMM,Zu2013TCOMM}.

By applying the MMSE matrix inversion to $\tilde {\bm H} \tilde {\bm A}$, we have
\begin{equation} \label{eqn:h_dag}
\begin{split}
  \tilde{\bm H}^{\dag} =  \tilde{\bm A}^{\text H}\tilde{\bm H}^{\text H}(\tilde{\bm H}\tilde{\bm A}\tilde{\bm A}^{\text H}\tilde{\bm H}^{\text H} + \rho \bm I_{2M})^{-1} \triangleq [\tilde{\bm H}_1^{\dag}, \cdots, \tilde{\bm H}_K^{\dag}],
  \end{split}
\end{equation}
where $\tilde {\bm H}_k^{\dag} \in \mathbb R^{2N \times 2M_k}$. Then we perform the QR decomposition to get $\tilde {\bm H}_k^{\dag} = \tilde {\bm F}_k \tilde{\bm R_k}$,
 where $\tilde {\bm F}_k \in \mathbb R^{2N \times 2M_k}$ is an orthogonal matrix and $\tilde{\bm R_k}  \in \mathbb R^{2M_k \times 2M_k}$ is an upper triangular matrix. {\color{red} The first component of the precoder is thus chosen as $\tilde {\bm P}_{k1} = \tilde {\bm F}_k$.} Once $\tilde{\bm P}_{k1}$ is ready, the effective channel matrix can be obtained as in \refeqn{eqn:h_k_e}. The matrices $\tilde{\bm P}_{k2}$ and $\tilde{\bm G}_k$ can be then calculated through SVD of $\tilde{\bm H}_{\text{e}_k}$.  This precoding method is referred to as WL-S-GMI.

A summary of the proposed widely-linear BD-type precoding algorithms is given in Table \ref{tab:wl_bd}.  The main difference among WL-BD, WL-RBD and WL-S-GMI is the way to calculate $\tilde{\bm P}_{k1}$, which is summarized in Table \ref{tab:pk1}.

\vspace*{-1em}
\begin{table}[!htp]
\caption{Proposed Widely-Linear BD-Type Precoding Algorithms.}
\vspace*{-3em}
\label{tab:wl_bd}
\small
\begin{center}
\begin{tabular}{ll}
\hline
\hline
\textbf{Steps} & \textbf{Operations} \\
\hline
 1 &  Obtain $\tilde{\bm H}\tilde{\bm A}$ by channel estimation;  \\
 2 &  {\color{red} $\tilde {\bm H}_{-k}\tilde{\bm A} = [(\tilde{\bm H}_1\tilde{\bm A})^{\text T}, \cdots, (\tilde{\bm H}_{k-1}\tilde{\bm A})^{\text T}, (\tilde{\bm H}_{k+1}\tilde{\bm A})^{\text T}, \cdots, (\tilde{\bm H}_K\tilde{\bm A})^{\text T}]^{\text T}$;}  \\
 3 &  For $k=1,\ldots, K$: \\
 \, 3.1 & \quad Calculate $\tilde{\bm P}_{k1}$ according to Table \ref{tab:pk1};\\
 \, 3.2 &  \quad $\tilde {\bm H}_{\text e _k} = \tilde {\bm H}_k \tilde{\bm A} \tilde{\bm P}_{k1}$; \\
 \, 3.3 &   \quad Perform SVD to get  $\tilde {\bm H}_{\text e _k} = \tilde {\bm U}_{k} \tilde{\bm \Sigma}_k \tilde{\bm V}_k^{\text H}$; \\
 \, 3.4 & \quad $\tilde{\bm P}_{k2} = \tilde{\bm V}_k$ and {\color{red} $\tilde{\bm G}_k = \tilde {\bm U}_{k}^{\text H}$};  \\
 \, 3.5 & \quad Select the power loading matrix $\tilde{ \bm \Gamma}_k$; \\
 4 & $\bm P_{\text a} = [\tilde{\bm P}_{11} \tilde{\bm P}_{12} \tilde{ \bm \Gamma}_1, \ldots, \tilde{\bm P}_{K1} \tilde{\bm P}_{K2} \tilde{ \bm \Gamma}_K ]$;  \\
 5 & { $\lambda_{\text{WL-BD}} = 2 P_{\text T} / \mathbb E \{ {\text {Tr}} [  {\bm P}_{\text a} {\bm P}_{\text{a}}^{\text T} ]\}$;}  \\
 6 & $\tilde{\bm P} = \sqrt{\lambda_{\text{WL-BD}}} \bm P_{\text a}$;  \\
 7 & {\color{red} The receive filter matrix $\tilde {\bm G} = \text{diag}\{\tilde {\bm G}_1, \ldots, \tilde {\bm G}_K \}$;} \\
 8 & The received signal is $\tilde {\bm G} \tilde {\bm y} = \tilde {\bm G} (\tilde{\bm H} \tilde {\bm A} \tilde{\bm P} \tilde{\bm s} + \tilde {\bm n})$. \\
\hline
\end{tabular}
\end{center}
\end{table}
\vspace*{-3em}

Note that the power loading schemes can be either water-filling for maximizing the sum rate, or equal power loading, or based on the improved diversity precoding approach in \cite{Stankovic2008TWC}. A detailed discussion is beyond the scope of this paper. We will simply assume equal power allocation in the following analysis.

\begin{table*}[!htp]
\caption{Methods to Calculate $\tilde{\bm P}_{k1}$.}
\vspace*{-3em}
\label{tab:pk1}
\begin{center}
\begin{tabular}{lcl}
\hline
\hline
\textbf{Algorithm} & \textbf{Steps} & \textbf{Operations} \\
\hline
\textbf{WL-BD} & 1 &   Perform SVD to get $\tilde{\bm H}_{-k}\tilde{\bm A} =\tilde{\bm U}_{-k} \tilde {\bm \Sigma}_{-k} [\tilde {\bm V}_{-k1}, \tilde{\bm V}_{-k0} ]^{\text H}$  \\
                & 2 &  $\tilde{\bm P}_{k1} = \tilde{\bm V}_{-k0}$  \\

 \textbf{WL-RBD} & 1 & Perform SVD to get  $\tilde{\bm H}_{-k}\tilde{\bm A} =\tilde{\bm U}_{-k} \tilde {\bm \Sigma}_{-k} \tilde{\bm V}_{-k}^{\text H}$ \\
                 & 2 & $\tilde{\bm P}_{k1} = \tilde{\bm V}_{-k} \left ( \tilde {\bm \varSigma}_{-k}^{\text H} \tilde {\bm \varSigma}_{-k} + \rho \bm I_{2N} \right)^{-\frac{1}{2}}$  \\

\textbf{WL-S-GMI} & 1 & Calculate$ \tilde {\bm H}_k^{\dag}$ according to \refeqn{eqn:h_dag}  \\
                    & 2 & Apply QR decomposition to get $\tilde {\bm H}_k^{\dag} = \tilde {\bm F}_k \tilde{\bm R_k}$ \\
                    & 3 & $\tilde {\bm P}_{k1} = \tilde {\bm F}_k$ \\
\hline
\end{tabular}
\end{center}
\end{table*}
\vspace*{-3em}

{
\emph{Remark:} When taking IQI into account, the precoding matrices designed using the real-valued signal model generally do not satisfy \refeqn{eqn:transform} and thus can not be represented in equivalent complex-valued matrices. However, the real-valued symbol vector after precoding can be inversely transformed into an equivalent complex-valued symbol vector.
}

\vspace*{-1em}
\section{Performance Analysis}

{
In order to show more insights on the proposed widely-linear precoding schemes, in this section the performance of WL-ZF and WL-BD precoding is analyzed in terms of sum rates, multiplexing gain, power offset and computational complexity. To facilitate the analysis, we adopt an affine approximation of the sum data rate developed in \cite{Shamai2001TIT}.

\begin{definition}[\cite{Shamai2001TIT}] \label{def:app}
The sum data rate is well approximated by
  $C(P_{\text T}) = S^{\infty} (\log_2 P_{\text T} - L^{\infty}) + {\it o} (1)$,
where $S^{\infty}$ is the multiplexing gain and $L^{\infty}$ is the power offset which are defined, respectively, as:{\color{red}
\vspace*{-1em}
\begin{equation} \label{eqn:def_mg}
\begin{split}
  S^{\infty} \triangleq \lim \limits_{P_{\text T}\to \infty} \frac{C(P_{\text T})}{\log_2 (P_{\text T})},  \quad
  L^{\infty} \triangleq \lim \limits_{P_{\text T}\to \infty} \left [ \log_2 (P_{\text T}) - \frac{C(P_{\text T})}{S^{\infty}}  \right ].
  \end{split}
\end{equation}}
\end{definition}
\vspace*{-1em}

We will use this tool to derive the multiplexing gain and power offset of WL-ZF and WL-BD in the following subsections.

\vspace*{-1em}
\subsection{Comparison between WL-ZF and ZF}

In order to analyze the performance of WL-ZF, we compare it with ZF in \cite{Peel2005TCOM} and assume perfect IQ branches for ZF  unless otherwise specified. The precoding matrix of ZF is given by
  $\bm P_{\text{ZF}} = \sqrt{\lambda_{\text{ZF}}} \bm H^{\text H} (\bm H \bm H^{\text H})^{-1}$,
where the power normalization factor is defined as
 \begin{equation}\label{eqn:lambda_zf}
   \lambda_{\text{ZF}} = \frac{P_{\text T}}{\mathbb E \{ \text{Tr} [\bm P_{\text{ZF}} \bm P^{\text H}_{\text{ZF}}]\}} = \frac{P_{\text T}}{\mathbb E \{\text{Tr} [(\bm H \bm H^{\text H})^{-1}]\}}.
 \end{equation}

The sum rate of ZF is given by
\begin{equation} \label{eqn:sr_zf}
  C_{\text {ZF}} = \sum \limits_{k=1}^K \log_2(1+\text{SINR}_{\text{ZF},k})= K \log_2\left (  1 + \frac{1}{\sigma_n^2} \lambda_{\text{ZF}}  \right ),
\end{equation}
where $\text{SINR}_{\text{ZF},k}$ represents the received signal-to-interference-plus-noise ratio (SINR) at user $k$. According to Definition \ref{def:app}, the multiplexing gain and power offset of ZF are given by
  $S^{\infty}_{\text{ZF}} = K, \quad L^{\infty}_{\text{ZF}} = \log_2\sigma_n^2 + \log_2\left[ \mathbb E \left \{ \text{Tr} [ (\bm H \bm H^{\text T})^{-1} ] \right ]\right \}$.

{\color{red} The difference between WL-ZF and ZF are fourfold: 1) The signal dimension is doubled from $K$ to $2K$; 2) Since WL-ZF transmits only real-valued signals, the data rate on each parallel sub-channel is halved; 3) The power normalization factor becomes $\lambda_{\text {WL-ZF}}$; 4) Both the transmit power and the noise variance for each sub-channel are halved.} The sum rate of WL-ZF is thus given by
\vspace*{-1em}
\begin{equation} \label{eqn:sr_wl_zf}
\begin{split}
  C_{\text{WL-ZF}} = \sum \limits_{k=1}^{2K} \frac{1}{2} \log_2(1+\text{SINR}_{\text{WL-ZF},k})
                   &= 2K \times \frac{1}{2} \log_2 \left (  1 + \frac{1}{\sigma_n^2} \lambda_{\text{WL-ZF}}   \right ) \\
                   & = K \log_2\left(  1 + \frac{1}{\sigma_n^2} \lambda_{\text{WL-ZF}} \right),
  \end{split}
\end{equation}
where $\text{SINR}_{\text{WL-ZF},k}$ represents the received SINR at user $k$ for WL-ZF. The multiplexing gain and power offset of WL-ZF are given by
\begin{equation} \label{eqn:s_l_wl_zf}
  S^{\infty}_{\text{WL-ZF}} = K, \quad L^{\infty}_{\text{WL-ZF}}  = \log_2 \sigma_n^2 + \log_2 \left[ \frac{1}{2} \mathbb E \left \{\text{Tr} [ (\tilde{\bm H} \tilde{\bm A} \tilde{\bm A}^{\text T} \tilde {\bm H}^{\text T})^{-1} ] \right \} \right ].
\end{equation}



We summarize the comparison  between ZF and WL-ZF in Theorem \ref{th:wl_zf}.

\begin{theorem} \label{th:wl_zf}
When the transmitter does not have IQI, i.e., $\bm A_1 = \bm I$ and $\bm A_2 = \bm 0$, WL-ZF has the same multiplexing gain and power offset as ZF. However, when the transmitter has IQI:
\begin{enumerate}
  \item WL-ZF has the same multiplexing gain as that of ZF with ideal IQ branches. The achieved multiplexing gain equals the number of users, i.e.,
        $S_{\text{WL-ZF}}^{\infty} = S_{\text{ZF}}^{\infty} = K$.
  \item Denote $\Delta \triangleq L_{\text{WL-ZF}}^{\infty} - L_{\text{ZF}}^{\infty}$ as the power offset loss of WL-ZF compared to ZF with ideal IQ branches. Assuming that: 1) $\theta_1, \ldots,\theta_N$ are i.i.d with zero-mean and variance $\sigma_{\theta}^2$; 2) $g_1,\ldots, g_N$ are i.i.d with zero-mean and variance $\sigma_g^2$; 3) The expectations in $L_{\text{WL-ZF}}^{\infty}$ are taken over $\bm H$, $\theta_1, \ldots,\theta_N$ and $g_1,\ldots, g_N$, then we have
\begin{equation} \label{eqn:power_loss_1}
    \Delta \approx \log_2\left [ 1 + (\sigma_{\theta}^2 + 4\sigma_g^2)\frac{K+1}{N+1} \right ],
\end{equation}
  which is simplified by denoting $\beta \triangleq \frac{K}{N}$ when $K$ and $N$ are large and $\sigma_{\theta}^2$ is small, as
  \begin{equation} \label{eqn:power_loss_2}
    \Delta \approx \log_2\left [ 1 + 4\sigma_g^2 \beta \right ].
\end{equation}
\end{enumerate}
\end{theorem}

\begin{IEEEproof}
See Appendix \ref{apd:th_wl_zf}.
\end{IEEEproof}

Theorem \ref{th:wl_zf} shows that compared with ZF with perfect IQ branches, WL-ZF in a system with IQI has no multiplexing gain loss, while the power offset loss of WL-ZF is determined by the IQ parameters and the system scale, i.e., the ratio of $K$ to $N$. Note that in large-scale systems, $\beta$ is usually small and thus the power offset loss of WL-ZF is limited. Therefore, WL-ZF will approach the performance of ZF without IQI. }



\vspace*{-1em}
\subsection{Comparison between WL-BD and BD}

In this subsection, we compare the performance of WL-BD in the presence of IQI with that of BD under perfect IQ branches.

The sum rate of WL-BD in the downlink is calculated as $C_{\text{WL-BD}} = \sum_{k=1}^K R_{\text{WL-BD}, k}$, where $R_{\text{WL-BD},k}$ is the data rate of the $k$-th user. {\color{red} Let $\tilde {\bm G}_k$ be the receive filter of the $k$-th user, and then multiplying the received signal vector by  $\tilde {\bm G}_k$ yields}
\begin{equation*}
\begin{split}
  \tilde {\bm d}_k = \tilde{\bm G}_k \tilde{\bm y}_k
                   = \tilde {\bm G}_k \tilde{\bm H}_k \tilde {\bm A} \tilde{\bm P}_k \tilde{\bm s}_k + \tilde {\bm G}_k \tilde{\bm H}_k \tilde {\bm A} \tilde{\bm P}_{-k} \tilde{\bm s}_{-k} + \tilde{\bm G}_k \tilde{\bm n}_k
                    \triangleq \tilde {\bm Q} _k \tilde {\bm s}_k + \tilde {\bm Q} _{-k} \tilde {\bm s}_{-k} + \tilde{\bm G}_k \tilde {\bm n}_k,
\end{split}
\end{equation*}
where
$\tilde{\bm P}_{-k} = [\tilde{\bm P}_1 ,\cdots, \tilde{\bm P}_{k-1}, \tilde{\bm P}_{k+1}, \cdots \tilde{\bm P}_K],
\tilde{\bm s}_{-k} = [\tilde{\bm s}_1^{\text T} ,\cdots, \tilde{\bm s}_{k-1}^{\text T}, \tilde{\bm s}_{k+1}^{\text T}, \cdots, \tilde{\bm s}_K^{\text T}]^{\text T}$,
and $\tilde{\bm Q}_k = \tilde {\bm G}_k \tilde{\bm H}_k \tilde {\bm A} \tilde{\bm P}_k$, $\tilde{\bm Q}_{-k} = \tilde {\bm G}_k \tilde{\bm H}_k \tilde {\bm A} \tilde{\bm P}_{-k}$.  Assuming Gaussian signaling is used, the data rate of the $k$-th user is thus given by
\begin{equation*}
  R_{\text{WL-BD},k} = \frac{1}{2} \log_2 \left \{ \frac{  \det [ \tilde{\bm Q}_k \tilde{\bm Q}_k^{\text T}  + \tilde{\bm Q}_{-k} \tilde{\bm Q}_{-k}^{\text T} + \sigma_n^2 \tilde{\bm G}_k \tilde{\bm G}_k^{\text T}] }{ \det [\tilde{\bm Q}_{-k} \tilde{\bm Q}_{-k}^{\text T} + \sigma_n^2 \tilde{\bm G}_{k} \tilde{\bm G}_{k}^{\text T} ] }  \right \}.
\end{equation*}

For WL-BD precoding, $\tilde{\bm Q}_{-k} = \bm 0$ and $\tilde{\bm G}_k$ does not affect the data rates. Therefore, for WL-BD we have
\vspace*{-1em}
\begin{equation*}
  R_{\text{WL-BD},k} = \frac{1}{2} \log_2\det \left [  \bm I_{2M_k}  + \frac{1}{\sigma^2_n}\tilde{\bm H}_k \tilde{\bm A} \tilde{\bm P}_k \tilde{\bm P}_k^{\text T} \tilde{\bm A}^{\text T} \tilde{\bm H}_k^{\text T}   \right ].
\end{equation*}

The following analysis is based on two assumptions:
\begin{itemize}
  \item [-] AS1: {\color{red} $D_k = 2M_k = 2L_k$, i.e., the data streams of each user are fully used and the number of data streams is twice the number of receive antennas.}
  \item [-] AS2: Equal power allocation is used across all the data streams,  i.e., $\tilde {\bm \Gamma}_k = \sqrt{\frac{P_{\text T}}{M}}\bm I_{2L_k}$, $k=1,\cdots,K$.  Note that $\mathbb E\{\tilde {\bm s}_k \tilde{\bm s}_k^{\text T}\} = \frac{1}{2} \bm I_{2L_k}$ for the real-valued signal model.
\end{itemize}

According to AS1 and AS2, the data rate of the $k$-th user can be expressed as:
\begin{equation} \label{eqn:rk_as}
  R_{\text{WL-BD},k} = \frac{1}{2} \log_2\det  \left [  \bm I_{2M_k}  + \frac{\lambda_{\text{WL-BD}}P_{\text T}}{M\sigma^2_n}\tilde{\bm H}_k \tilde{\bm A} \tilde{\bm P}_{k1} \tilde{\bm P}_{k1}^{\text T} \tilde{\bm A}^{\text T} \tilde{\bm H}_k^{\text T}   \right ].
\end{equation}

\begin{proposition} \label{pro:dr_no_iqi}
When the transmitter does not have IQI, i.e., $\bm A_1 = \bm I$ and $\bm A_2 = \bm 0$, WL-BD achieves the same data rate as BD,  which is given by
\begin{equation} \label{eqn:rk_bd}
  R_{\text{BD},k} = \log_2\det \left ( \bm I_{M_k}  + \frac{P_{\text T}}{M\sigma^2_n} {\bm H_k} \bm V_{-k0} \bm V_{-k0}^{\text H} {\bm H}_k^{\text H}  \right ),
\end{equation}
for $k=1,\cdots, K$, where $\bm V_{-k0}$ contains the right singular vectors corresponding to zero singular values of $\bm H_{-k} = [\bm H_1^{\text T}, \cdots, \bm H_{k-1}^{\text T}, \bm H_{k+1}^{\text T},\cdots, \bm H_K^{\text T}]^{\text T}$.
\end{proposition}
\begin{IEEEproof}
See Appendix \ref{apd:pro1}.
\end{IEEEproof}

There is no performance loss introduced by widely-linear precoding in terms of data rates when there is no IQI. However, when IQI does exist, WL-BD has significantly improved performance and approaches that of BD with ideal IQ branches, as shown in the following proposition.

\begin{proposition} \label{pro:mul_high}
When the transmitter has IQI, WL-BD has the same multiplexing gain as BD in the absence of IQI.
\end{proposition}
\begin{IEEEproof}
{\color{red} The $k$-th user's data rate of BD is given by \refeqn{eqn:rk_bd}, and we have
  $S_{\text{BD},k}^{\infty} = M_k, \quad L_{\text{BD},k}^{\infty}=\log_2\sigma_n^2+\log_2 M-\frac{1}{M_k}\log_2\det[\bm H_k \bm P_k\bm P_k^{\text T}\bm H_k^{\text H}]$,
where $\bm P_k$ which substitutes $\bm V_{-k0}$ is the BD precoding matrix for the $k$-th user. According to Definition \ref{def:app}, \refeqn{eqn:rk_bd} is well approximated in the high SNR region as
\begin{equation*}
  R_{\text{BD},k} \cong M_k \log_2 \frac{P_{\text T}}{\sigma^2} - M_k \log_2 M + \log_2 \det (\bm H_k \bm P_k \bm P_k^{\text H} \bm H_k^{\text H}).
\end{equation*}
}  Therefore, the sum data rate of BD without IQI is described as
\begin{equation*}
\begin{split}
  C_{\text{ BD}} = \sum \nolimits_{k=1}^K R_{\text {BD}, k} & \cong \sum \nolimits_{k=1}^K \log_2 \det (\bm H_k \bm P_k \bm P_k^{\text H} \bm H_k^{\text H})
                + M \log_2 \frac{P_{\text T}}{\sigma^2} - M \log_2 M.
\end{split}
\end{equation*}
The same results can also be found in \cite{Lee2007TIT}. According to the definition of the multiplexing gain in \refeqn{eqn:def_mg}, the multiplexing gain of BD is $M$, the total number of the receive antennas.

Similarly to BD, we have
\begin{equation*}
\begin{split}
  C_{\text {WL-BD}}
  = \sum \nolimits_{k=1}^K R_{\text {WL-BD}, k}
  \cong& \frac{1}{2} \sum \nolimits_{k=1}^K  \left ( 2M_k \log_2 \frac{\lambda_{\text{WL-BD}}P_{\text T}}{\sigma^2} - 2M_k \log_2 M \right ) + J \\
   \cong& M\log_2 \left( \frac{P_{\text T}}{\sigma^2} \right ) - M \log_2 \left(\frac{M}{\lambda_{\text{WL-BD}}} \right) + J.
\end{split}
\end{equation*}
where $J = \frac{1}{2} \sum_{k=1}^K \log_2 \det (\tilde {\bm H}_k \tilde {\bm A}\tilde{\bm P}_{k1} \tilde{\bm P}_{k1}^{\text T} \tilde {\bm A} ^{\text T}\tilde{\bm H}_k^{\text T})$. The multiplexing gain is easy to compute according to \refeqn{eqn:def_mg} and is given by $M$, which is the same as BD.
\end{IEEEproof}

Although there is no multiplexing gain loss for WL-BD, the power offset is different from that of BD without IQI. It comes from the value of $\lambda_{\text{WL-BD}}$ and the choices of the precoding matrices, which are related to the IQ parameters. In large-scale MIMO systems, this power offset will converge to some constant almost surely.  However, its mathematical expression is difficult to obtain and we leave it for future work.

\vspace*{-1em}
\subsection{Computational Complexity}

{
Since the inverted matrices of both ZF and WL-ZF have the same dimension (for real-valued elements), the computational complexity of the two are of the same order. Therefore, we omit the complexity analysis of WL-ZF. For similar reasons, WL-MF and WL-MMSE also have similar complexity with their linear counterparts.}

In terms of widely-linear BD-type precoders, we use the total number of floating-point operations (FLOPs) involved in the algorithm to study its computational complexity.  Each real-valued multiplication or addition counts for 1 FLOP, while one complex-valued multiplication and addition counts for 6 FLOPs and 2 FLOPs, respectively.  The total number of FLOPs of some basic matrix operations are summarized as follows:
\begin{itemize}
 \item The addition of two $N \times K$ real matrix requires $NK$ FLOPs, while that of complex matrices is $2NK$;
  \item The multiplication of an $N \times K$ and a $K \times M$ real matrix requires $NM(2K-1)$ FLOPs, while that of complex matrices is $NM(8K-2)$;
  \item The inverse of a $N \times N$ real matrix requires $\frac{4}{3}N^3$;
  \item For QR decomposition of an $M\times N$ ($M \ge N$) real matrix, the required number of FLOPs is $4(M^2N-MN^2+N^3/3)$;
  \item The FLOPs required by SVD of an $K \times M$ ($K\le M$) complex-valued matrix is the same as that of an $2K \times 2M$ real-valued matrix\cite{Zu2012LCOMM}. When only $\bm \Sigma$ and $\bm V$ are obtained, the number of FLOPs is $32KM^2 + 104K^3$,   and when $\bm \Sigma$, $\bm V$ and $\bm U$ are obtained, it requires  $32M^2K + 176K^3$\cite{Golub1996MC}.
\end{itemize}

Note that the real and imaginary components of a complex-valued scalar are stored separately in the hardware. The $\mathcal T$-transform actually requires only twice the memory space, but does not increase the computational complexity. Therefore, we will exclude it in the analysis. In the following, we also assume for simplicity that all the users have the same number of antennas, i.e., $M_1 = M_2 = \ldots =M_k =\ldots = M_K$.

For WL-BD, to calculate the SVD of $\tilde{\bm H}_{-k}\tilde{\bm A}$ requires $N_1 = 32M_{-k}N^2 + 104M_{-k}^3$. Similarly, a matrix product and an SVD are involved in computing $\tilde {\bm H}_{\text e _k} = \tilde {\bm U}_{k} \tilde{\bm \Sigma}_k \tilde{\bm V}_k^{\text H}$, which yields $N_2 = 4M_k^2(4N-1)$ and $ N_3 = 208M_k^3$ FLOPs, respectively. {\color{red} Note that although an SVD is required for computing both $\tilde{\bm P}_{k1}$ and $\tilde{\bm P}_{k2}$, the complexity of the latter is much lower}. Compared with WL-BD, WL-RBD demands an extra matrix product to calculate $\tilde{\bm P}_{k1}$, which accounts for $4N^2$. For WL-S-GMI, we need to compute two matrix products and a matrix inverse in \refeqn{eqn:h_dag}, which requires $N_4 = 4M^2(4N-1)+4N^2(4M-1) + \frac{32}{3}N^3$. For each user, WL-S-GMI involves an SVD and a QR decomposition, which accounts for $N_3$ and $N_5 = 32M_kN^2 -8NM_k^2+\frac{4}{3}M_k^2$ FLOPs. The total number of FLOPs required by the three algorithms are summarized in Table \ref{tab:bd_com}.

\vspace*{-1em}
\begin{table}[!htb]
\color{red}
\caption{Computational Complexity of Proposed
Widely-Linear BD-Type Precoding Schemes where $\gamma\triangleq \frac{M}{N}$ and assuming $K$ and $N$ are large.}
\vspace*{-3em}
\label{tab:bd_com}
\small
\begin{center}
\begin{tabular}{cc}
\hline
\hline
\textbf{ Algorithms} & \textbf{Number of FLOPs} \\
\hline
\textbf{WL-BD} & $\gamma K N^3[(104+\frac{208}{K^3})\gamma^2+\frac{16}{K^2}\gamma+32]$ \\
\textbf{WL-RBD} & $\gamma K N^3[(104+\frac{208}{K^3})\gamma^2+\frac{16}{K^2}\gamma+32]+4N^2$  \\
\textbf{WL-S-GMI} & $N^3[\frac{208}{K^2}\gamma^3+(16-\frac{8}{K})\gamma^2+48\gamma+\frac{32}{3}]+\frac{4\gamma^2}{3K}N^2$ \\
\hline
\end{tabular}
\end{center}
\end{table}
\vspace*{-3em}



 {\color{red} From Table \ref{tab:bd_com}, the complexity increase of WL-BD compared with BD is rather small, i.e., below $\frac{1}{8\gamma KN}$ and could be considered negligible when the scale of the system is large.}


\vspace*{-1em}
\subsection{Implementation Aspects}

In order to implement the proposed widely-linear algorithms, an estimate of $\tilde{\bm H} \tilde{\bm A}$ is required. This could be done through channel estimation approaches based on \refeqn{eqn:rx_sig_all}. According to \refeqn{eqn:rx_sig_all}, another solution to the IQI is constructing $\tilde{\bm P} = \tilde{\bm A}^{-1} \tilde{\bm P}_0$. This compensates for the IQI, which requires the estimates of both $\tilde{\bm H}$ and $\tilde{\bm A}$, and thus will increase the training and estimation complexity. Unlike the compensation based schemes, WL-BD and WL-ZF do not need respective information of $\tilde{\bm H}$ and $\tilde{\bm A}$, but only their matrix product. Therefore, they are simple to implement.

\vspace*{-1em}
\section{Numerical Results}

In this section, we evaluate the performance of the proposed widely-linear precoding schemes through simulations. We compare the proposed widely-linear precoding schemes with their linear counterparts, e.g., MF  \cite{Lo1999TCOM} \& MMSE \cite{Peel2005TCOM}, RBD  \cite{Stankovic2008TWC} \& S-GMI \cite{Zu2013TCOMM} for single-antenna and multiple-antenna users, respectively.

{  Unless otherwise defined herein, in all the simulations there are $K=20$ users, each equipped with $M_k=2$ antennas for multiple-antenna scenarios, while the number of transmit antennas at BS is set to $N=100$. In terms of IQ parameters, we consider ``SETUP 1'' where $\sigma_g^2 = 0.1, \sigma_{\theta}^2 = 0.003$ as the {\it default configuration}. SETUP 1 is the case when $g_n$ and $\theta_n$ have standard deviations of $0.33$ and $3^\circ$, respectively, which are in the typical range of the IQ parameters\cite{Tarighat2007TWC}. In comparison, ``SETUP 0'' with $\sigma_g^2 = 0.05, \sigma_{\theta}^2 = 0.001$ and ``SETUP 2'' where $\sigma_g^2 = 0.2, \sigma_{\theta}^2 = 0.01$ are also considered for light and severe IQI, respectively. The SNR in all the figures is defined as $\frac{P_{\text T}}{\sigma_n^2}$. The bit error rates (BER) are evaluated considering that {\color{red} quadrature phase shift keying (QPSK)} modulation are used at the transmitter. {\color{red} For linear precoding schemes under IQI, $\bm H\bm A_1$ is used as the channel estimate to calculate the precoders.} All the simulations are averaged over 10000 channel realizations.}

  \vspace*{-1em}
\begin{figure}[!thp]
\begin{minipage}[t]{0.45\linewidth}
\centering
\includegraphics[width=\columnwidth]{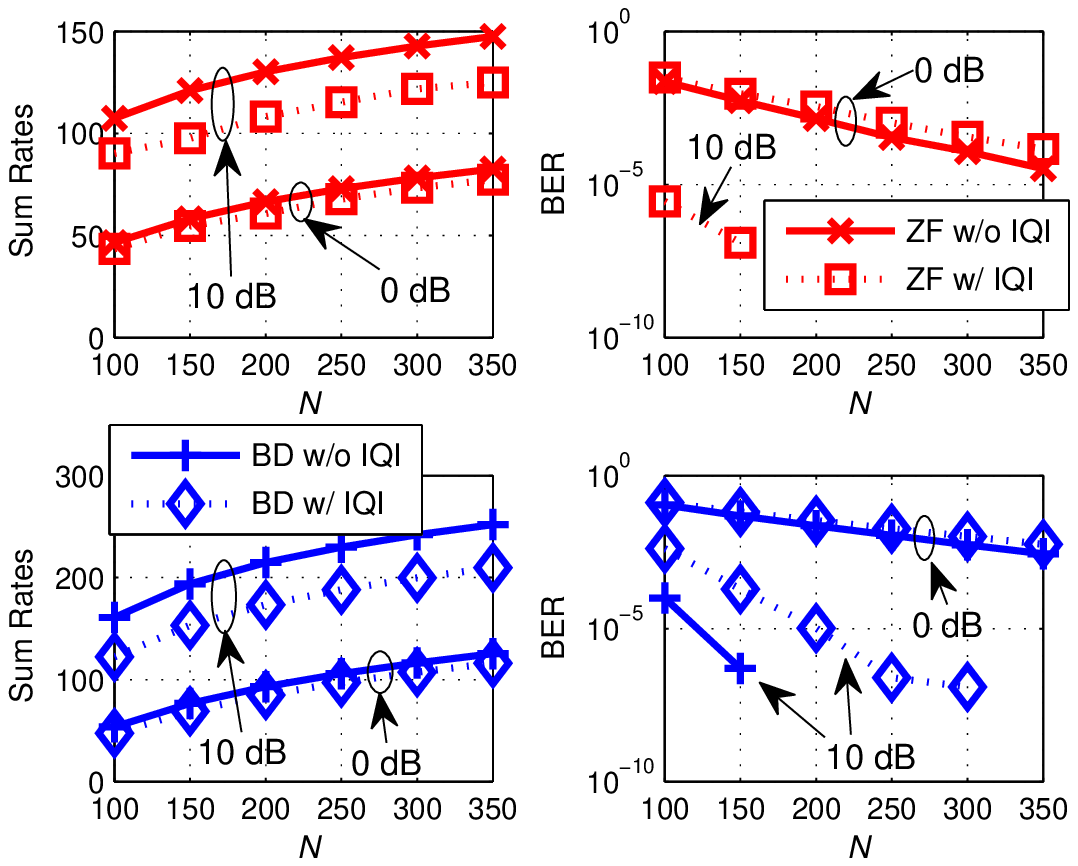}\\
  \vspace*{-2em}
  \caption{\color{red} Performance loss of ZF (the top two) and BD (the bottom two) with IQI with respect to $N$ in terms of sum rates (bits/channel use) and BER. The SNR is 0, 10 dB. There are 20 single antenna users and 20 two-antenna users for ZF and BD, respectively.}\label{fig:n_not_help}
\end{minipage}
\begin{minipage}[t]{0.45\linewidth}
\centering
\includegraphics[width=\columnwidth]{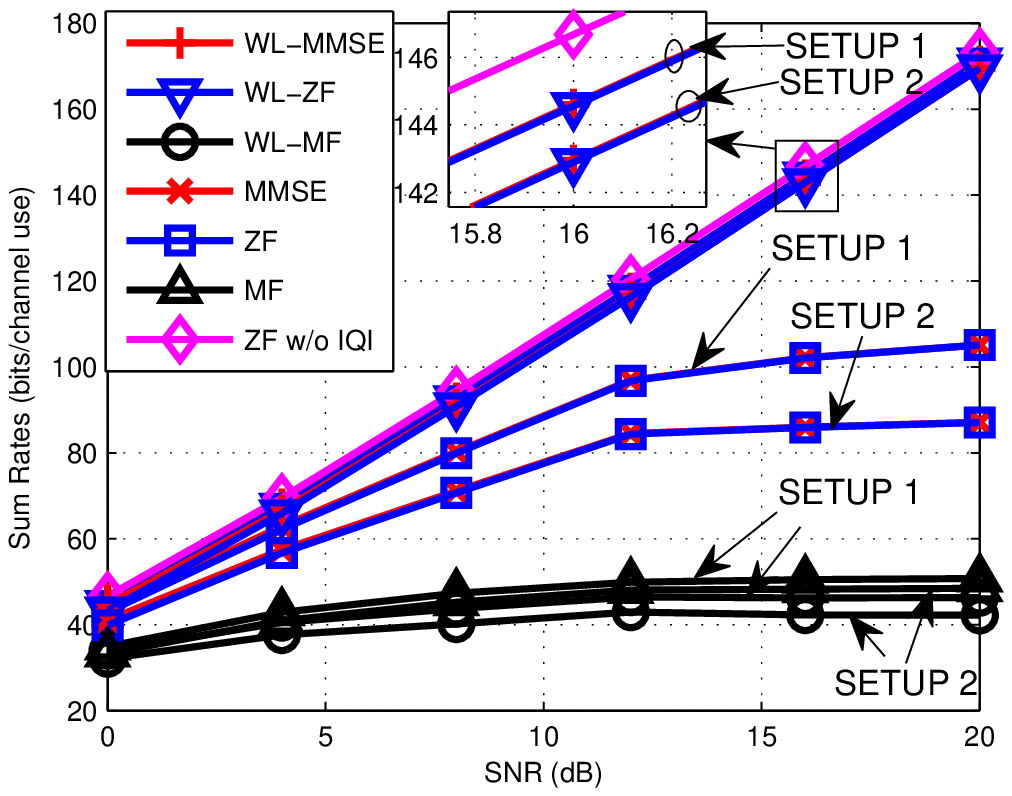}\\
  \vspace*{-2em}
  \caption{Sum rates of ZF \& MMSE in \cite{Peel2005TCOM}, MF in \cite{Lo1999TCOM} and their widely-linear counterparts under IQI.}\label{fig:sr_zf}
\end{minipage}
\end{figure}
\vspace*{-2em}

{
\reffig{fig:n_not_help} shows the performance loss of ZF and BD when IQI is considered at the transmitter. Unlike noise and independent inter-user interference which generally diminish with large $N$\cite{Rusek2013MSP}, it can be seen that when IQI exists the performance loss of both ZF and BD in terms of sum rates and BER does not vanish with respect to $N$. Moreover, this IQI-originated performance loss becomes large for high SNR, e.g., when SNR is 10 dB, ZF and BD lose 20\% of their sum rates. Therefore, one has to take IQI into account for downlink design.
}

{
For scenarios with single-antenna users, \reffig{fig:sr_zf} shows the sum rates of ZF \& MMSE in \cite{Peel2005TCOM}, MF in \cite{Lo1999TCOM} and their widely-linear counterparts under IQI. The performance of both WL-ZF(WL-MMSE, WL-MF) and ZF(MMSE, MF) degrade when IQI becomes more severe.
However, WL-ZF and WL-MMSE outperform ZF and MMSE significantly, especially in the high SNR region. The sum rates of ZF and MMSE level out in the high SNR region as a result of the IQI. In contrast, the proposed WL-ZF and WL-MMSE can efficiently suppress the negative impact of IQI and approaches that of ZF without IQI. In the high SNR region, WL-ZF has the same diversity gain as ZF (i.e., the same slope of the curves) with a minor power offset (i.e., the shift of the curves) around $10\log_{10}( 1+4\beta\sigma_g^2 ) = 0.3$ dB for $\sigma_g^2 = 0.1$ and 0.6 dB for $\sigma_g^2 = 0.2$,  which verifies the results in Theorem \ref{th:wl_zf}.

 In \reffig{fig:sr_zf}, WL-MF performs worse than MF. The reason is that WL-MF deals with twice the number of sub-channels as that for MF. Since WL-MF (MF) offers no inter-stream interference control and aims to maximize the SNR other than the SINR at the receiver, it could increase the interference level at the receiver and thus degrades the performance.
}

\vspace*{-1em}
\begin{figure}[!thp]
\begin{minipage}[t]{0.45\linewidth}
\centering
 \includegraphics[width=\columnwidth]{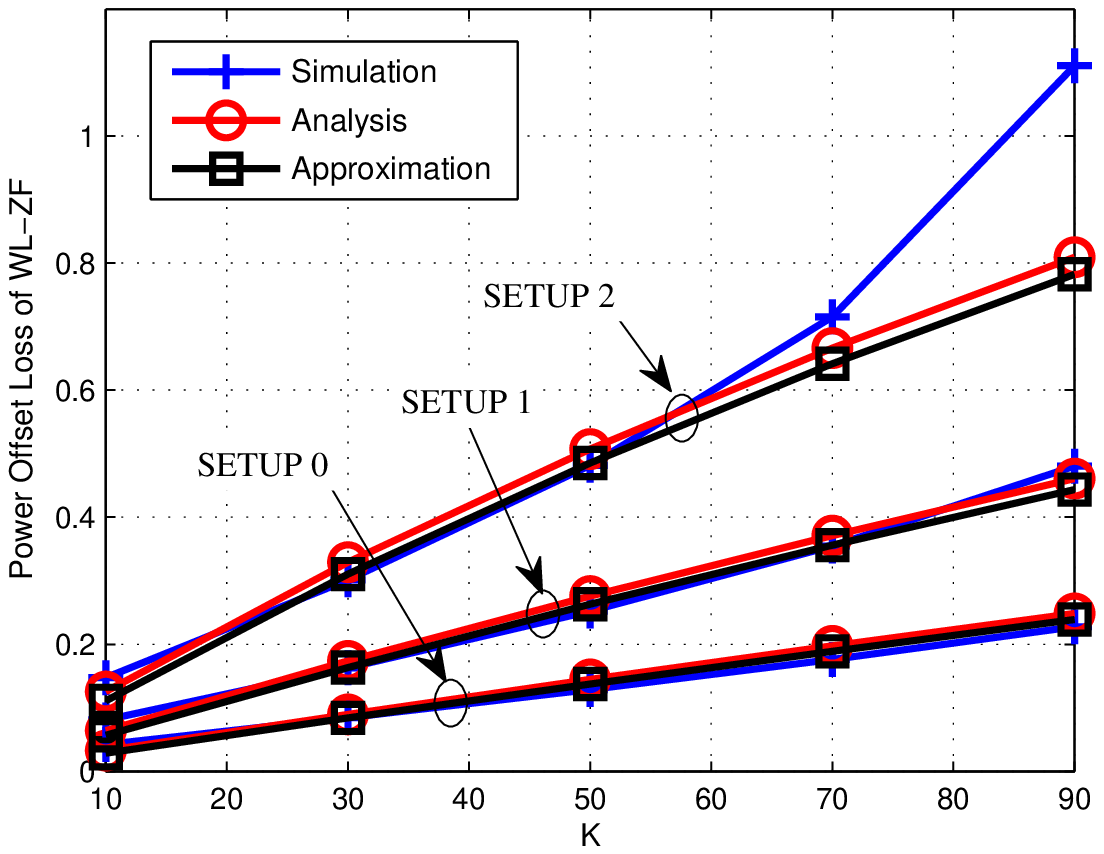}\\
  \vspace*{-2em}
  \caption{Power offset loss of WL-ZF when compared with ZF with perfect IQ branches.}\label{fig:ratio}
\end{minipage}
\begin{minipage}[t]{0.45\linewidth}
\centering
\includegraphics[width=\columnwidth]{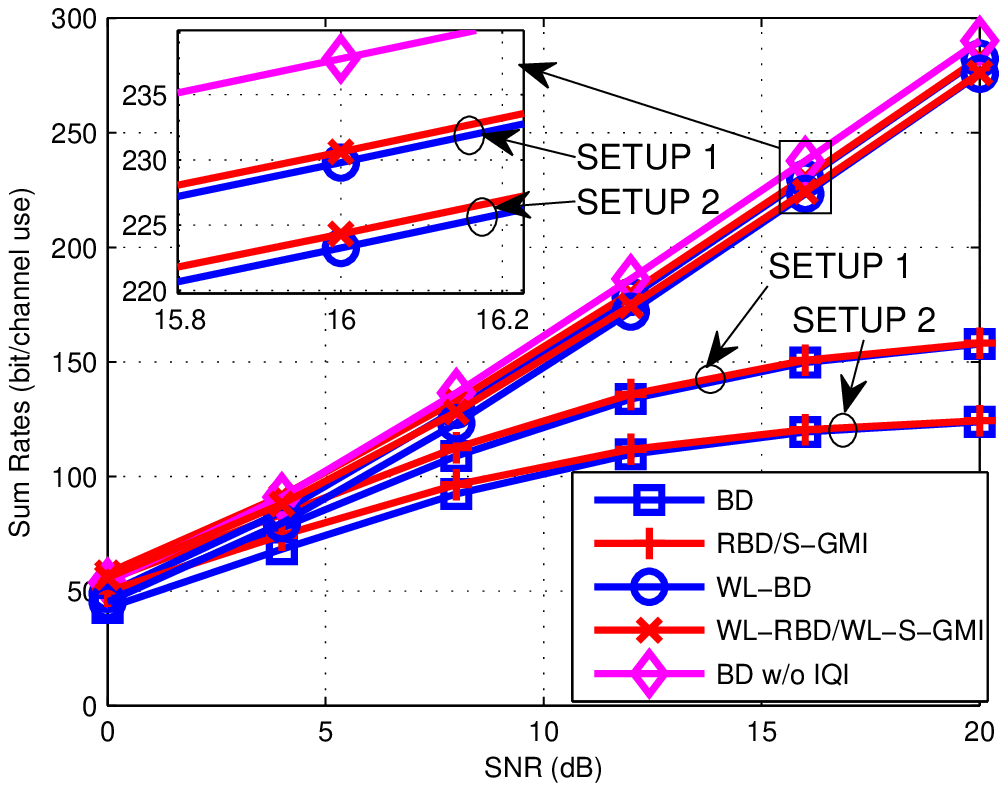}\\
  \vspace*{-2em}
  \caption{Sum rates of BD \cite{Spencer2004TSP,Choi2004TWC}, RBD \cite{Sung2009TCOMM}, S-GMI \cite{Zu2013TCOMM}, the proposed WL-BD, WL-RBD and WL-S-GMI under IQI.}\label{fig:sr_bd}
\end{minipage}
\end{figure}
\vspace*{-2em}




{
\reffig{fig:ratio} shows the power offset loss of WL-ZF compared with ZF with ideal IQ branches when $N=100$. The analysis results are obtained using \refeqn{eqn:power_loss_1} and the approximation is made according to \refeqn{eqn:power_loss_2}. The simulation results show that the analysis is very accurate for most cases. The approximation results also give precise prediction of the power offset loss. As $\beta$ {\color{red} increases}, the power offset loss of WL-ZF gets larger. For `SETUP 2' which indicates very severe IQI, the analytical results are not accurate for $\beta>0.7$. {\color{red} This inaccuracy comes from the Taylor expansion. Improved accuracy could be achieved by using higher order expansions. However, the analysis becomes complicated. In fact, $\beta$ is usually smaller than $0.5$ in large-scale MIMO systems in order to take advantage of the excess degrees of freedom. Moreover, for the typical IQ imbalance parameters, i.e., 'SETUP 1', Theorem \ref{th:wl_zf} is accurate enough.}
}


{
\reffig{fig:sr_bd} shows the sum rates comparison of BD \cite{Spencer2004TSP,Choi2004TWC}, RBD \cite{Sung2009TCOMM}, S-GMI \cite{Zu2013TCOMM}, their widely-linear conterparts WL-BD, WL-RBD and WL-S-GMI under IQI, where the curves for RBD (WL-RBD) and S-GMI (WL-S-GMI) coincide with each other. The widely-linear precoding schemes significantly outperform the original schemes in the high SNR region where IQI is the key factor. It is interesting to see the performance of BD, RBD and S-GMI levels out when the SNR is high.  In contrast, the widely-linear approaches are able to tackle the IQI and show much better sum rates performance. There is a slight performance gap between WL-BD and BD without IQI. However, when there is no IQI for WL-BD, it will achieve the same sum-rates as BD, as proved in Proposition \ref{pro:dr_no_iqi}.
}
%
%
%

\vspace*{-1em}
\begin{figure}[!htb]
  \centering
  \includegraphics[width=2.8in]{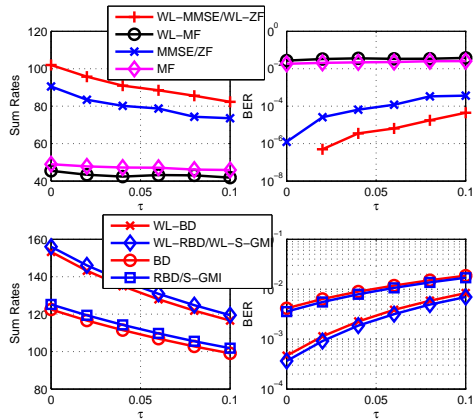}\\
  \vspace*{-2em}
  \caption{\color{red} Impact on sum rates (bits/channel use) and BER of imperfect CSI for single-antenna users (the top two) and multiple-antenna users (the bottom two). The SNR value is 10 dB. Note that the curves of ZF and MMSE, WL-ZF and WL-MMSE, S-GMI and RBD, WL-S-GMI and WL-RBD coincide with each other.}\label{fig:imp}
\end{figure}
\vspace*{-2em}

{\color{red}
In order to show the performance of the proposed precoding schemes with imperfect channel state information, we introduce the channel estimation error model in \cite{Zu2013TCOMM} for the linear precoders, which is ${\bm H}_{\text e} = {\bm H} {\bm A_1} + \sqrt{\tau}\bm N_{\text e}$
where the entries of $\bm N_{\text e}$ are i.i.d Gaussian with zero-mean and unit variance. Here $\tau$ is a parameter to control the channel estimation accuracy with a larger value indicating a more severe estimation error. Similarly, for the widely-linear precoders we have $\tilde{\bm H}_{\text e} = \tilde{\bm H} \tilde{\bm A} + \sqrt{\tau}\tilde{\bm N}_{\text e}$
where the entries of $\tilde{\bm N}_{\text e}$ are i.i.d Gaussian with zero-mean and the variance is 0.5 since real-valued signals are considered.

\reffig{fig:imp} shows the performance of the proposed algorithms under different levels of channel estimation error. With increased $\tau$, the performance of WL-ZF, WL-MMSE, WL-BD, WL-RBD and WL-S-GMI degrades, while WL-MF is more robust. However, according to (25) in \cite{Biguesh2004ICC}, the value of $\tau$ with $N=100$, $K=20$ and SNR=$10$ dB is usually below 0.01 when MMSE channel estimation is used. Although the proposed schemes degrade with increased channel estimation error, they outperform their linear counterparts significantly.
}

\vspace*{-1.5em}
\section{Conclusion}
\vspace*{-1em}
{
In this paper, widely-linear  precoding schemes have been proposed based on a real-valued signal model to deal with IQI in the large-scale MIMO downlink. The analysis shows that WL-ZF and WL-BD achieve the same sum data rates as ZF and BD if the transmitter does not present IQI. Furthermore, when there exists IQI at the transmitter, WL-ZF and WL-BD have the same multiplexing gain as ZF and BD with perfect IQ branches, which equals $M$, the total number of receive antennas. Moreover, we have proved that there is a minor power offset loss for WL-ZF, which is related to the system scale and the IQ parameters. Numerical results have verified the analysis and shown that the widely-linear precoders significantly outperform conventional precoders in the presence of IQI.
}


\appendices

\vspace*{-2em}
\section{Selection of the Power Normalization Factor}
\label{app:norm}
{\color{red}
We take $N=1$ as an example to illustrate the selection of the power normalization factor. The extension to general cases is straightforward.

The average transmit power is normalized by introducing a normalization factor $\mu_0$ such that
 $z = \mu_0(a_1 x + a_2 x^*)$,
where $x$ is the transmitted signal and $z$ is the signal degraded by IQ imbalance, and the IQ imbalance parameters are given by
  $a_1 = \cos(\theta/2) + \text{j}g \sin(\theta/2),   a_2 = g\cos(\theta/2) - \text{j} \sin(\theta/2)$.
{Applying the $\mathcal T$-transform}, we get $\mathcal{T}(z) = \mu_0[\mathcal{T}(a_1) + \mathcal{T}(a_2)\bm{E}_1 ] \mathcal{T} (x)$.
Denote $\tilde{\bm  z} = \mathcal{T}(z)$, $\tilde{\bm  A}_1 = \mathcal{T}(a_1)$, $\tilde{\bm  A}_2 = \mathcal{T}(a_2)$ and $\tilde{\bm  x} = \mathcal{T}(x)$. we have $\tilde{\bm  z} = [\tilde{\bm  A}_1 + \tilde{\bm  A}_2\bm E_1] \tilde{\bm  x} \triangleq \tilde{\bm A}\tilde{\bm x}$.

To normalized the transmit power such that $\mathbb E \{\|\tilde{\bm z}\|_{\text F}^2\} = \mathbb E\{\|\tilde{\bm x}\|_{\text F}^2\}$, we have
\begin{equation}\label{eqn:expct_eq}
  \mathbb E\{\|\tilde{\bm z}\|_{\text F}^2\} = \mathbb E \left \{ \text{Tr}\left [\mu_0^2 \tilde{\bm x} \tilde{\bm x}^{\text T} \tilde{\bm A}^{\text T} \tilde{\bm A}  \right ] \right \} = \mathbb E \left \{ \text{Tr}\left [ \mu_0^2 \tilde{\bm x} \tilde{\bm x}^{\text T} \mathbb E \{\tilde{\bm A}^{\text T} \tilde{\bm A} \}  \right ] \right \},
\end{equation}
in which
\vspace*{-1em}
\begin{equation*}
   \tilde{\bm A}^{\text T} \tilde{\bm A}  =\begin{bmatrix}
    1+g^2+2g\cos\theta & -g^2\sin\theta  \\
    -g^2 \sin \theta & 1+g^2-2g\cos\theta
   \end{bmatrix}.
\end{equation*}
Since we assume that $\theta \sim U(0,\sigma_{\theta}^2)$, $g \sim \mathcal N(0, \sigma_{g}^2)$ and $\theta$, $g$ are independent, we have $\mathbb E \{\tilde{\bm A}^{\text T} \tilde{\bm A} \} = (1+\sigma_{g}^2) \bm I_2$. Therefore, $\mu_0$ is obtained from \refeqn{eqn:expct_eq} and given by
   $\mu_0 = \frac{1}{\sqrt{1+\sigma_g^2}}$.
}

\vspace*{-1em}
\section{Proof of Theorem \ref{th:wl_zf}}
\label{apd:th_wl_zf}

In order to prove Theorem \ref{th:wl_zf}, two useful lemmas are given first, which give results on the expectations of products formed by moments of entries of a Haar matrix.

{
\begin{lemma}[Lemma 1.1, \cite{Hiai2000ASM}] \label{lem:haar_corr_0}
{\color{red} Denote $\mathbb N_0$, $\mathbb N$ as the set of non-negative integers and positive integers, respectively. Let $\bm U = [u_{ij}]_{N \times N}$ be a Haar matrix. Let $l \in \mathbb N$, and $i_1, \ldots, i_l$, $j_1,\ldots,j_l \in \{1,\ldots, N\}$ be the subscript indexes. Denote $k_1, \ldots, k_l$, $m_1, \ldots, m_l \in \mathbb N_0$. If $\exists i\in \{1,\ldots, N\}$ which satisfies $\sum_{r\in\{r|i_r = i\}}(k_r - m_r) \neq 0$, or  $\exists j \in \{1,\ldots, N\}$ which satisfies $\sum_{r\in \{r|j_r = j\}}(k_r - m_r) \neq 0$, then we have
  $\mathbb E \{ [u_{i_1j_1}^{k_1} (u_{i_1j_1}^*)^{m_1}] \times \cdots \times [u_{i_lj_l}^{k_l} (u_{i_lj_l}^*)^{m_l}] \} = 0$.
}
\end{lemma}

Lemma \ref{lem:haar_corr_0} shows that if there exists $u_{i_rj_r}$, the power of which  is different from that of its complex conjugate, the expectation above is always 0.

\begin{lemma}[Proposition 1.2, \cite{Hiai2000ASM}]  \label{lem:haar_corr_1}
If $1\le i,j,i',j'\le N$, $i\neq i'$, $j \neq j'$, and $\bm U = [u_{ij}]_{N \times N}$ is a Haar matrix, we have
\begin{equation*}
  \begin{array}{ll}
      (1)\ \mathbb E\{ |u_{ij}|^2\}  = \frac{1}{N},    & (2)\ \mathbb E\{ |u_{ij}|^4\}  = \frac{2}{N(N+1)}, \\
      (3)\ \mathbb E\{ |u_{ij}|^2|u_{i'j}|^2\} =\mathbb E\{ |u_{ij}|^2|u_{ij'}|^2\} = \frac{1}{N(N+1)},   & (4)\ \mathbb E\{|u_{ij}|^2|u_{i'j'}|^2\}  = \frac{1}{N^2-1}, \\
      (5)\ \mathbb E\{u_{ij}u_{i'j'}u_{ij'}^*u_{i'j}^*\}  = -\frac{1}{N(N^2-1)}. &
   \end{array}
\end{equation*}
\end{lemma}

It is easy to prove that $\tilde{\bm H} \tilde{\bm A} = \tilde {\bm H}$ when no IQI is present at the transmitter. Thus the precoding matrix of WL-ZF is exactly equivalent to that of ZF. It is also straightforward to obtain $S_{\text{WL-ZF}}^{\infty} = S_{\text{ZF}}^{\infty} = K$, therefore we omit the detailed proof.

To prove the result on power offset loss of WL-ZF, we need to compare $\frac{1}{2} \mathbb E \{ \text{Tr} [ (\tilde{\bm H} \tilde{\bm A} \tilde{\bm A}^{\text T} \tilde {\bm H}^{\text T})^{-1} ]\}$ and $\mathbb E \{ \text{Tr} [ (\bm H \bm H^{\text H})^{-1} ]\}$. According to \refeqn{eqn:t_her} in Corollary \ref{lem:t_trans_2}, we have $\mathbb E \{ \text{Tr} [ (\tilde{\bm H} \tilde{\bm H}^{\text T})^{-1} ] \} = 2\mathbb E \{\text{Tr} [ (\bm H \bm H^{\text H})^{-1} ]\}$. Therefore, we only need to compare $\mathbb E \{ \text{Tr} [ (\tilde{\bm H} \tilde{\bm H}^{\text T})^{-1} ]\}$ and $\mathbb E \{\text{Tr} [ (\tilde{\bm H} \tilde{\bm A} \tilde{\bm A}^{\text T} \tilde {\bm H}^{\text T})^{-1} ]\}$.


%

{\color{red} Note that
\begin{equation}
  \tilde{\bm A} \tilde{\bm A}^{\text T} = \frac{1}{1+\sigma_g^2} \bm G \bm \varTheta \bm G
\end{equation}}
where $\bm G = \text{diag} \{ 1+g_1,\ldots,1+g_N,1-g_1,\ldots,1-g_N \}$ and
\begin{equation}\label{eqn:Theta}
  \bm \varTheta = \begin{bmatrix}
    \bm I_N & \text{diag}\{ -\sin(\theta_1),\ldots,-\sin(\theta_N) \} \\
    \text{diag}\{ -\sin(\theta_1),\ldots,-\sin(\theta_N) \}  & \bm I_N
  \end{bmatrix}
\end{equation}
Therefore, we have
\begin{equation}
  \begin{split}
  \mathbb E \{ \text{Tr} [ (\tilde{\bm H} \tilde{\bm A} \tilde{\bm A}^{\text T} \tilde {\bm H}^{\text T})^{-1} ] \}
  =  (1+\sigma_g^2) \mathbb E \{ \text{Tr} [ (\tilde{\bm H} \bm G \bm \varTheta \bm G \tilde {\bm H}^{\text T})^{-1} ]\}.
  \end{split}
\end{equation}

To compare $\mathbb E \{ \text{Tr} [ (\tilde{\bm H} \tilde{\bm H}^{\text T})^{-1} ]\}$ and $ (1+\sigma_g^2)\mathbb E \{\text{Tr} [ (\tilde{\bm H} \bm G \bm \varTheta \bm G   \tilde {\bm H}^{\text T})^{-1} ] \}$, we observe that $\bm G$ and $\bm \varTheta$ have a small deviation from the identity matrix. Therefore, we can analyze the derivatives of
$$
f(\sigma_g, \bm G, \bm \varTheta) = (1+\sigma_g^2)\mathbb E \{ \text{Tr} [ (\tilde{\bm H} \bm G \bm \varTheta \bm G  \tilde {\bm H}^{\text T})^{-1} ]\},
$$
with respect to $\bm G$, $\bm \varTheta$ and $\sigma_g$.

The strategy of the following proof is first to consider when
$g_1 = g_2 = \ldots = g_N = g_0$ and $\theta_1 = \theta_2 = \ldots = \theta_N = \theta_0$, where $g_0$ and $\theta_0$ are zero-mean distributed with {\color{red}variances} given by $\sigma_g^2$ and $\sigma_{\theta}^2$, respectively.  Then we analyze the derivatives of $f(\sigma_g, \bm G, \bm \varTheta) = f(\sigma_g, g_0,\theta_0)$ assuming the expectation is taken over $\bm H$ only. After that, we take the expectation in $f(\sigma_g, \bm G, \bm \varTheta)$ over $g_0$ and $\theta_0$. Finally, the results are extended to general cases when $g_1, g_2, \ldots, g_N$ and $\theta_1, \theta_2, \ldots, \theta_N$ are i.i.d, respectively.

Let us start with the second-order Taylor expansion of $f(\sigma_g, g_0,\theta_0)$, which is given by\cite{TaylorOnline}
\begin{equation} \label{eqn:taylor_f}
  \begin{split}
    f(\sigma_g, g_0,\theta_0) & = f(0,0,0) + \left ( \sigma_g \frac{\partial }{\partial \sigma_g}  + g_0 \frac{\partial }{\partial g_0} + \theta_0 \frac{\partial }{\partial \theta_0} \right ) \left. f \right |_{(0,0,0)}  \\
    & + \frac{1}{2}  \left (  \sigma_g \frac{\partial }{\partial \sigma_g} + g_0  \frac{\partial }{\partial g_0} + \theta_0 \frac{\partial }{\partial \theta_0}  \right )^2 \left. f\right |_{(0,0,0)} + o(\sigma_g^2 + g_0^2 + \theta_0^2),
  \end{split}
\end{equation}
in which all the partial derivatives are evaluated at point $(0,0,0)$. Note that $f(0,0,0) = \mathbb E \{ \text{Tr} [ $ $(\tilde{\bm H} \tilde{\bm H}^{\text T})^{-1} ]\}$ is related to the power offset in the case without IQI.

We need to compute the partial derivatives in \refeqn{eqn:taylor_f}, which are summarized in Lemma \ref{lem:partial_sum}. Due to the symmetry of second derivatives, i.e., $\frac{\partial^2 f}{\partial g_0 \partial \theta_0} = \frac{\partial^2 f}{\partial \theta_0 \partial g_0 }$ and etc, the order of partial derivatives with respect to different variables does not matter.

\begin{lemma}\label{lem:partial_sum}
The partial derivatives in \refeqn{eqn:taylor_f} are given by
{\color{red} \begin{subequations}\label{eqn:partial_sum}
  \begin{align}
     \left. \frac{\partial f }{\partial \sigma_g} \right |_{(0,0,0)} &= 0,  \left. \frac{\partial f }{\partial g_0} \right |_{(0,0,0)} = 0,  \left. \frac{\partial f }{\partial \theta_0} \right |_{(0,0,0)} = 0,  \\
     \left. \frac{\partial^2 f }{\partial \sigma_g \partial g_0 } \right |_{(0,0,0)} &= 0, \left. \frac{\partial^2 f }{\partial \sigma_g \partial \theta_0 } \right |_{(0,0,0)} = 0, \left. \frac{\partial^2 f }{\partial \theta_0 \partial g_0 } \right |_{(0,0,0)} = 0, \\
     \left. \frac{\partial^2 f }{\partial \sigma_g^2 } \right |_{(0,0,0)} &= 4\mathbb E \{\text{Tr} [({\bm H}  {\bm H}^{\text H})^{-1}]\}, \\
     \left. \frac{\partial^2 f }{\partial g_0^2 } \right |_{(0,0,0)} &= 16 \mathbb E \left \{\text{Tr} \left [  \bm H^{\text H} ({\bm H}  {\bm H}^{\text H})^{-2}  {\bm H} [\bm H^{\text H} ({\bm H}  {\bm H}^{\text H})^{-1}  {\bm H}]^* \right ] \right \} -  4\mathbb E \{\text{Tr} [({\bm H}  {\bm H}^{\text H})^{-1}]\},\\
     \left. \frac{\partial^2 f }{\partial \theta_0^2 } \right |_{(0,0,0)} &= 4 \mathbb E \left \{ \text{Tr} \left [  \bm H^{\text H} ({\bm H}  {\bm H}^{\text H})^{-2}  {\bm H} [\bm H^{\text H} ({\bm H}  {\bm H}^{\text H})^{-1}  {\bm H}]^* \right ]\right \}.
  \end{align}
\end{subequations}}
in which
\begin{equation} \label{eqn:drv_1st}
  \mathbb E \{ \text{Tr} [(\bm H \bm H^{\text H})^{-1}]\} = \frac{K}{N-K},
\end{equation}
\begin{equation} \label{eqn:drv_2st_1}
  \mathbb E \left \{ \text{Tr} \left [  \bm H^{\text H} ({\bm H}  {\bm H}^{\text H})^{-2}  {\bm H} [\bm H^{\text H} ({\bm H}  {\bm H}^{\text H})^{-1}  {\bm H}]^* \right ]\right \} = \frac{K^2+K}{(N-K)(N+1)}.
\end{equation}
\end{lemma}
\begin{IEEEproof}
Let us first prove \refeqn{eqn:partial_sum}. Since $\frac{\partial f }{\partial \sigma_g} = 2\sigma_g\mathbb E \{\text{Tr} [ (\tilde{\bm H} \bm G \bm \varTheta \bm G  \tilde {\bm H}^{\text T})^{-1} ]\}$, it is straightforward to get
\begin{equation}\label{eqn:partial_sigma_g}
\begin{split}
  \left. \frac{\partial f }{\partial \sigma_g} \right |_{(0,0,0)} &= 0, \left. \frac{\partial^2 f }{\partial \sigma_g^2} \right |_{(0,0,0)} =  2\mathbb E \{\text{Tr} [ (\tilde{\bm H} \tilde {\bm H}^{\text T})^{-1}]\} = 4\mathbb E \{\text{Tr} [ ({\bm H} {\bm H}^{\text T})^{-1}]\}, \\
  \left. \frac{\partial^2 f }{\partial \sigma_g \partial g_0 } \right |_{(0,0,0)} &= \left. \frac{\partial^2 f }{\partial \sigma_g \partial \theta_0 } \right |_{(0,0,0)} = 0.
  \end{split}
\end{equation}

The first order partial derivative of $f(\sigma_g,g_0,\theta_0)$ with respect to $g_0$ is given by
\begin{equation} \label{eqn:1st_g0}
  \frac{\partial f}{ \partial  g_0} = - (1+\sigma_g^2) \mathbb E \left \{ \text{Tr} \left [  \bm D_1 \bm D_2 \bm D_1  \right ]\right \}.
\end{equation}
where $\bm D_1 = (\tilde{\bm H} \bm G \bm \varTheta \bm G  \tilde {\bm H}^{\text T})^{-1}$, and  $\bm D_2 = \frac{\partial (\bm D_1^{-1})}{\partial g_0} =\tilde {\bm H} [\bm E_N \bm \varTheta \bm G + \bm G \bm \varTheta \bm E_N ] \tilde{\bm H}^{\text T}$.
When $g_0 = 0, \theta_0 = 0$, we have $\bm G = \bm \varTheta = \bm I_{2N}$. Using \refeqn{eqn:t_her} gives
\begin{equation} \label{eqn:dg=0} 
  \left. \frac{\partial f}{ \partial  g_0} \right|_{(0,0,0)} = - 2 \mathbb E \left \{\text{Tr} \left [   \tilde{\bm H}^{\text T} (\tilde{\bm H} \tilde {\bm H}^{\text T})^{-2} \tilde {\bm H} \bm E_N  \right ]\right \} =  0.
\end{equation}

Taking the derivative of  \refeqn{eqn:1st_g0} with respect to $g_0$ yields
\begin{equation}\label{eqn:2nd_g0}
\begin{split}
  \left. \frac{\partial^2 f}{ \partial  g_0^2} \right |_{(0,0,0)}  &= - \left. 2 \mathbb E \left \{ \text{Tr} \left [ \frac{\partial \bm D_1}{ \partial  g_0}\bm D_2 \bm D_1 \right ]\right \}\right |_{(0,0,0)} - \left. \mathbb E \left \{\text{Tr} \left [ \bm D_1 \frac{\partial \bm D_2}{ \partial  g_0} \bm D_1 \right ] \right \} \right  |_{(0,0,0)}  \\
  &=  8 \mathbb E \{ \text{Tr} [ \tilde{\bm H}^{\text T} (\tilde{\bm H}  \tilde {\bm H}^{\text T})^{-2} \tilde {\bm H} \bm E_N \tilde{\bm H}^{\text T} (\tilde{\bm H}  \tilde {\bm H}^{\text T})^{-1} \tilde {\bm H} \bm E_N ] \} - 2\mathbb E \{ \text{Tr} [(\tilde{\bm H} \tilde {\bm H}^{\text T})^{-1}]\}  \\
  & \mathop = \limits^{(a)} 16 \mathbb E \left  \{ \text{Tr} \left [  \bm H^{\text H} ({\bm H}  {\bm H}^{\text H})^{-2}  {\bm H} [\bm H^{\text H} ({\bm H}  {\bm H}^{\text H})^{-1}  {\bm H}]^* \right ] \right \}-  4\mathbb E \{ \text{Tr} [({\bm H}  {\bm H}^{\text H})^{-1}] \}.
  \end{split}
\end{equation}
where $(a)$ follows Corollary \ref{lem:t_trans_2}.

The first order partial derivative of $f(\sigma_g,g_0,\theta_0)$ with respect to $\theta_0$ is given by
\begin{equation}\label{eqn:1st_theta0}
  \frac{\partial f}{ \partial  \theta_0} = - (1+\sigma_g^2) \mathbb E \{ \text{Tr} [  \bm D_1 \bm D_3  \bm D_1 ]\},
\end{equation}
where $\bm D_3 = \frac{\partial (\bm D_1^{-1})}{\partial \theta_0} = -\cos\theta_0 \tilde{\bm H} \bm G \bar{\bm I}_N \bm G \tilde{\bm H}^{\text T}$. When $\sigma_g=0, g_0 = 0, \theta_0 = 0$, using \refeqn{eqn:t_her} yields
\begin{equation}\label{eqn:1st_theta0_2}
\begin{split}
  \left. \frac{\partial f}{ \partial  \theta_0} \right |_{(0,0,0)} &= \mathbb E \left \{\text{Tr} \left [  \tilde{\bm H}^{\text T}(\tilde{\bm H}\tilde{\bm H}^{\text T})^{-2}\tilde{\bm H} \bar{\bm I}_N \right ]\right \} = 0
  \end{split}
\end{equation}

Taking the derivativeof  \refeqn{eqn:1st_theta0} with respect to $\theta_0$ yields
\begin{equation}\label{eqn:2nd_theta0}
\begin{split}
  \left. \frac{\partial^2 f}{ \partial \theta_0^2} \right |_{(0,0,0)}
  &= -\left [ \left. 2 \mathbb E \left \{\text{Tr} \left [ \frac{\partial \bm D_1}{ \partial  \theta_0}\bm D_3 \bm D_1 \right ]\right \}\right |_{(0,0,0)} + \left. \mathbb E \left \{\text{Tr} \left [ \bm D_1 \frac{\partial \bm D_3}{ \partial  \theta_0} \bm D_1 \right ] \right \} \right  |_{(0,0,0)} \right ] \\
  & \mathop = \limits^{(a)}  2 \mathbb E \{\text{Tr} [ \tilde{\bm H}^{\text T} (\tilde{\bm H}  \tilde {\bm H}^{\text T})^{-2} \tilde {\bm H} \bar{\bm I }_N \tilde{\bm H}^{\text T} (\tilde{\bm H}  \tilde {\bm H}^{\text T})^{-1} \tilde {\bm H} \bar {\bm I}_N ] \} \\
  & \mathop =  \limits^{(b)} 4 \mathbb E \left \{\text{Tr} \left [  \bm H^{\text H} ({\bm H}  {\bm H}^{\text H})^{-2}  {\bm H} [\bm H^{\text H} ({\bm H}  {\bm H}^{\text H})^{-1}  {\bm H}]^* \right ]\right \},
  \end{split}
\end{equation}
where $(a)$ follows that $\left. \frac{\partial \bm D_3}{\partial \theta_0} \right |_{(0,0,0)} = \bm 0_{K\times K}$ and $(b)$ is obtained using Corollary \ref{lem:t_trans_2}.

Now we have only $\left. \frac{\partial^2 f}{ \partial g_0 \partial \theta_0} \right |_{(0,0,0)} $ left. Taking the derivative of  \refeqn{eqn:1st_g0} with respect to $\theta_0$ gives
\begin{equation}\label{eqn:2nd_g0_theta0}
\begin{split}
  \left. \frac{\partial^2 f}{ \partial g_0 \partial \theta_0} \right |_{(0,0,0)}
  =& -\left [ \left. 2 \mathbb E \left \{\text{Tr} \left [ \frac{\partial \bm D_1}{ \partial  \theta_0}\bm D_2 \bm D_1 \right ]\right \}\right |_{(0,0,0)} + \left. \mathbb E \left \{\text{Tr} \left [ \bm D_1 \frac{\partial \bm D_2}{ \partial  \theta_0} \bm D_1 \right ] \right \} \right  |_{(0,0,0)} \right ] \\
  =& 4 \mathbb E \{\text{Tr} [ \tilde{\bm H}^{\text T} (\tilde{\bm H}  \tilde {\bm H}^{\text T})^{-2} \tilde {\bm H} \bar{\bm I }_N \tilde{\bm H}^{\text T} (\tilde{\bm H}  \tilde {\bm H}^{\text T})^{-1} \tilde {\bm H} \bm E_N ]\} \\
   & -2\mathbb E \{\text{Tr} [ \tilde{\bm H}^{\text T} (\tilde{\bm H}  \tilde {\bm H}^{\text T})^{-2} \tilde {\bm H} (\bm E_N \bar{\bm I}_N + \bar{\bm I_N}\bm E_N) ]\} \\
  \mathop = \limits^{(a)}&  8 \mathbb E \left \{\text{Tr} \left [ \text{Im} \{ \bm H^{\text H} ({\bm H}  {\bm H}^{\text H})^{-2}  {\bm H} [\bm H^{\text H} ({\bm H}  {\bm H}^{\text H})^{-1}  {\bm H}]^* \} \right ]\right \} \mathop = \limits^{(b)} 0,
  \end{split}
\end{equation}
where $(a)$ is obtained by using Lemma \ref{lem:t_trans}, \refeqn{eqn:en_in} in Corollary \ref{lem:t_trans_2} and $\bm E_N \bar{\bm I}_N + \bar{\bm I_N}\bm E_N = \bm 0_{2N\times 2N}$; $(b)$ follows that the trace of the product of two Hermitian matrices is real-valued, i.e., for two Hermitian matrices $\bm A$ and $\bm B$ it holds that $\text{Tr}^*[ \bm{AB}] = \text{Tr}^*[ (\bm{BA})^{\text H}] = \text{Tr}[ (\bm{BA})^{\text T} ] = \text{Tr}[ \bm{AB} ]$.

Combining \refeqn{eqn:partial_sigma_g}, \refeqn{eqn:dg=0}, \refeqn{eqn:2nd_g0}, \refeqn{eqn:1st_theta0_2}, \refeqn{eqn:2nd_theta0} and \refeqn{eqn:2nd_g0_theta0} results in \refeqn{eqn:partial_sum}.

Now let us calculate the expectations in \refeqn{eqn:partial_sum} which are taken over $\bm H$. Since $\bm H \bm H^{\text H}$ follows a Wishart distribution, \refeqn{eqn:drv_1st} is easily obtained using the property of Wishart distribution \cite{Tulino2004RMT}. To prove \refeqn{eqn:drv_2st_1}, applying SVD to $\bm H$ gives
\vspace*{-1em}
\begin{equation} \label{eqn:H_svd}
  \bm H = \bm U \bm {\varLambda}^{1/2} \bm V^{\text H},
\end{equation}
where $\bm U \in \mathbb C^{K \times K}$, $\bm V \in \mathbb C^{N \times K}$, $\bm {\varLambda} = \text{diag}\{ \lambda_1, \ldots, \lambda_K\}$, the diagonal entries of which are the eigenvalues of $\bm H \bm H^{\text H}$. Substituting \refeqn{eqn:H_svd} into \refeqn{eqn:partial_sum} yields
\begin{equation} \label{eqn:drv_2nd}
  \mathbb E \left \{\text{Tr} \left [  \bm H^{\text H} ({\bm H}  {\bm H}^{\text H})^{-2}  {\bm H} [\bm H^{\text H} ({\bm H}  {\bm H}^{\text H})^{-1}  {\bm H}]^* \right ] \right \}= \mathbb E \{\text{Tr} [  \bm V \bm {\varLambda}^{-1} \bm V^{\text H} \bm V^* \bm V^{\text T} ]\}.
\end{equation}
The columns of $\bm V$ are actually the $K$ eigenvectors with respect to the non-zero eigenvalues of $\bm H^{\text H}\bm H$. Because a Wishart matrix is unitary invariant, the matrix formed by its eigenvectors is a Haar matrix which is independent of the eigenvalues\cite{Tulino2004RMT}. Therefore,  $\bm V$ is independent of $\bm \varLambda$. Thus
\vspace*{-1em}
\begin{equation} \label{eqn:drv_2nd_3}
  \begin{split}
  \mathbb E \left \{ \text{Tr} \left [  \bm H^{\text H} ({\bm H}  {\bm H}^{\text H})^{-2}  {\bm H} [\bm H^{\text H} ({\bm H}  {\bm H}^{\text H})^{-1}  {\bm H}]^* \right ] \right \} & = \text{Tr} [ \mathbb E \{  \bm V^{\text H} \bm V^* \bm V^{\text T} \bm V \} \mathbb E \{\bm {\varLambda}^{-1} \} ] \\
  & \mathop = \limits^{(a)} \frac{1}{N-K} \mathbb E \{\text{Tr} [\bm V^{\text H} \bm V^* \bm V^{\text T} \bm V ]\}.
\end{split}
\end{equation}
where $(a)$ follows the mean of the inverse eigenvalues of a Wishart matrix in \cite{Jorswieck2001IS}. Denote $\bm V = [\bm v_1,\ldots, \bm v_K]$, where $\bm v_i = [v_{i1}, \ldots, v_{iN}]^{\text T}$, and we have
\begin{equation}
  \begin{split}
    \mathbb E \{\text{Tr} [\bm V^{\text H} \bm V^* \bm V^{\text T} \bm V ]\}  = \mathbb E \left \{\sum_{i=1}^K \sum_{j=1}^K \bm v_i^{\text H} \bm v_j^* \bm v_j^{\text T} \bm v_i  \right \}&= \mathbb E \left \{ \sum_{i=1}^K \sum_{j=1}^K \sum_{t=1}^N \sum_{l=1}^N v_{it}^*v_{jt}^*v_{il}v_{jl} \right \} \\
&=  \sum_{i=1}^K \sum_{j=1}^K \sum_{t=1}^N \sum_{l=1}^N \mathbb E \{ v_{it}^*v_{jt}^*v_{il}v_{jl} \}
  \end{split}.
\end{equation}
To derive $\mathbb E \{ v_{it}^*v_{jt}^*v_{il}v_{jl} \}$, consider the following four cases:
\begin{itemize}
  \item [a)] when $t\neq l$, $i\neq j$, according to Lemma \ref{lem:haar_corr_0}, $\mathbb E \{ v_{it}^*v_{jt}^*v_{il}v_{jl} \} = 0$£»
  \item [b)] when $t = l$, $i\neq j$, according to Lemma \ref{lem:haar_corr_1}, $\mathbb E \{ v_{it}^*v_{jt}^*v_{il}v_{jl} \} = \frac{1}{N(N+1)}$. This case has $(K^2-K)N$ terms in total;
  \item [c)] when $t \neq l$, $i= j$, according to Lemma \ref{lem:haar_corr_0}, $\mathbb E \{ v_{it}^*v_{jt}^*v_{il}v_{jl} \} = 0$£»
  \item [d)] when $t = l$, $i = j$, according to Lemma \ref{lem:haar_corr_1}, $\mathbb E \{ v_{it}^*v_{jt}^*v_{il}v_{jl} \} = \frac{2}{N(N+1)}$. This case has $KN$ terms in total.
\end{itemize}
To this end, it is easy to achieve
\begin{equation} \label{eqn:drv_2nd_2}
  \begin{split}
    \mathbb E \{\text{Tr} [\bm V^{\text H} \bm V^* \bm V^{\text T} \bm V ]\}  &=  \sum_{i=1}^K \sum_{j=1}^K \sum_{t=1}^N \sum_{l=1}^N \mathbb E \{ v_{it}^*v_{jt}^*v_{il}v_{jl} \}\\
&= (K^2-K)N\frac{1}{N(N+1)} + KN\frac{2}{N(N+1)} = \frac{K^2+K}{N+1}.
  \end{split}
\end{equation}
Substituting  \refeqn{eqn:drv_2nd_2} into \refeqn{eqn:drv_2nd_3} gives  \refeqn{eqn:drv_2st_1}.

The proof of Lemma \ref{lem:partial_sum} is completed.
\end{IEEEproof}

Substituting results in Lemma \ref{lem:partial_sum} into \refeqn{eqn:taylor_f} yields
\begin{equation} \label{eqn:taylor_f_2}
    f(\sigma_g, g_0,\theta_0) \approx f(0,0,0) + (2\theta_0^2 + 8g_0^2) \frac{K^2+K}{(N-K)(N+1)} + 2(\sigma_g^2 - g_0^2)\frac{K}{N-K}.
\end{equation}
Taking the expectation in $f(\sigma_g, g_0,\theta_0)$ over $g_0$ and $\theta_0$ gives
\begin{equation} \label{eqn:taylor_f_3}
    f(\sigma_g, g_0,\theta_0) \approx f(0,0,0) + (2\sigma_{\theta}^2 + 8\sigma_g^2) \frac{K^2+K}{(N-K)(N+1)}.
\end{equation}

Since $f(0,0,0) = \mathbb E \{ \text{Tr}[(\tilde {\bm H} \tilde{\bm H}^{\text H})^{-1}] \}= 2\mathbb E \{\text{Tr}[({\bm H} {\bm H}^{\text H})^{-1}] \}= \frac{2K}{N-K}$, from \refeqn{eqn:taylor_f_3} we have
\begin{equation} \label{eqn:power_loss_3}
    \Delta = L_{\text{WL-ZF}}^{\infty} - L_{\text{ZF}}^{\infty} \approx \log_2\left [ 1 + (\sigma_{\theta}^2 + 4\sigma_g^2)\frac{K+1}{N+1} \right ].
\end{equation}

The above results are extended to cases when $g_1,\ldots, g_N$ and $\theta_1, \ldots, \theta_N$ are i.i.d, respectively, by taking derivatives of $f(\sigma_g,\bm G,\bm \varTheta )$ with respect to each variable following a similar procedure.  Note that the expectation is taken over both $\bm H$ and the IQ parameters. Therefore, in the second-order Taylor expansion of $f(\sigma_g,\bm G,\bm \varTheta )$, only $f(0,\bm I, \bm I)$ and the terms related to $\frac{\partial^2 f}{\partial \sigma_g^2}$, $\frac{\partial^2 f}{\partial g_i^2}$ and $\frac{\partial^2 f}{\partial {\theta_i^2}}$ ($i=1,\ldots,N$) remain because the IQ parameters are i.i.d and have zero mean. $\frac{\partial^2 f}{\partial g_i^2}$ and $\frac{\partial^2 f}{\partial {\theta_i^2}}$ can be simply obtained by substituting $\bm E_N$ and $\bar{\bm I}$ in the derivation of $\frac{\partial^2 f}{\partial g_0^2}$ and $\frac{\partial^2 f}{\partial {\theta_0^2}}$ with $\bm E_N^{(i)}$ and $\bar{\bm I}^{(i)}$, respectively, where $\bm E_N^{(i)}$ and $\bar{\bm I}^{(i)}$ are formed by forcing all elements excluding the $i$-th and the $(N+i)$-th rows in $\bm E_N$ and $\bar{\bm I}$ to be 0. Finally, one achieves the same results as in \refeqn{eqn:power_loss_3}.
}

\vspace*{-1em}
\section{Proof of Proposition \ref{pro:dr_no_iqi}}
\label{apd:pro1}


{
It is easy to derive that $\lambda_{\text{WL-BD}}=1$ and thus the $k$-th user's data rates for WL-BD is calculated as
\vspace*{-1em}
\begin{equation}\label{eqn:proof_pro1_rk}
  R_{\text{WL-BD},k}= \frac{1}{2} \log_2\det \left [  \bm I_{2M_k}  + \frac{P_{\text T}}{M\sigma^2_n}\tilde{\bm H}_k \tilde{\bm P}_{k1} \tilde{\bm P}_{k1}^{\text T}  \tilde{\bm H}_k^{\text T}   \right ],
\end{equation}
where $\tilde{\bm P}_{k1}=\tilde{\bm V}_{-k0}$, and $\tilde{\bm H}_{-k}=\tilde{\bm U}_{-k} \tilde {\bm \Sigma}_{-k} [\tilde {\bm V}_{-k1}, \tilde{\bm V}_{-k0} ]^{\text H}$.

Let us frist write the SVD of $\mathcal T(\bm H_{-k})$ and $\bm H_{-k}$, respectively, as
\begin{equation}\label{eqn:svd_t_hk}
\begin{split}
  \mathcal T(\bm H_{-k}) = \bar{\bm U}_{-k} \bar {\bm \Sigma}_{-k} [\bar {\bm V}_{-k1}^{\rm H}, \bar{\bm V}_{-k0}^{\rm H} ], \quad
  \bm H_{-k} = \bm U_{-k} \bm \Sigma_{-k} [\bm V_{-k1}, \bm V_{-k0}]^{\text H},
\end{split}
\end{equation}
where $\bar {\bm V}_{-k1}^{\rm H}$ and $\bar{\bm V}_{-k0}^{\rm H} $, $\bm V_{-k1}$ and $\bm V_{-k0}$ contain the right singular vectors with respect to non-zero and zero singular values of $\mathcal T(\bm H_{-k})$ and $\bm H_{-k}$, respectively.

The essence of this proof is to show $\tilde{\bm P}_{k1} = \tilde{\bm V}_{-k0} = \mathcal T(\bm V_{-k0})$.  Since from \refeqn{eqn:H_k_c} we know ${\tilde {\bm H}}_{-k}$ is a permutation of rows of $\mathcal T(\bm H_{-k})$, according to Corollary \ref{lem:t_unitary}, we have $\tilde{\bm V}_{-k0}=\bar{\bm V}_{-k0}$. Therefore, we only need to prove $\bar{\bm V}_{-k0} = \mathcal T(\bm V_{-k0})$.

The $\mathcal T$-transform of $\bm H_{-k}$ is given by $ \mathcal T(\bm H_{-k}) = \mathcal T(\bm U_{-k}) \mathcal T(\bm \Sigma_{-k}) \mathcal T(\bm V_{-k}^{\text H})$. According to Corollary \ref{lem:t_unitary}, both $\mathcal T(\bm U_{-k})$ and $\mathcal T(\bm V_{-k})$ are orthogonal matrices. Since $\mathcal T(\bm \Sigma_{-k})$ is a diagonal matrix, $\mathcal T(\bm U_{-k}) \mathcal T(\bm \Sigma_{-k}) \mathcal T(\bm V_{-k}^{\text H})$ is actually the SVD of $\mathcal T(\bm H_{-k})$ with singular values arranged in a different order as that in \refeqn{eqn:svd_t_hk}.
}

Supposing $\bm \Sigma_{-k} = \text{diag} \{ \bm \Sigma_{-k1}, \bm 0 \}$, where $\bm \Sigma_{-k1}$ is a diagonal matrix with diagonal entries given by the non-zero singular values of $\bm H_{-k}$, we have $\mathcal T(\bm \Sigma_{-k}) = \text{diag}\{\bm \Sigma_{-k}, \bm \Sigma_{-k}\}$, and
\begin{equation*}
  \mathcal T(\bm V_{-k}) = \begin{bmatrix}  \text{Re}(\bm V_{-k1}) & \text{Re}(\bm V_{-k0}) & -\text{Im} (\bm V_{-k1}) & -\text{Im} (\bm V_{-k0})  \\
   \text{Im}(\bm V_{-k1}) & \text{Im}(\bm V_{-k0}) & \text{Re} (\bm V_{-k1}) & \text{Re} (\bm V_{-k0})
  \end{bmatrix}.
\end{equation*}
Note that since $ \mathcal T(\bm H_{-k}) = \mathcal T(\bm U_{-k}) \mathcal T(\bm \Sigma_{-k}) \mathcal T(\bm V_{-k}^{\text H}) = \bar{\bm U}_{-k} \bar {\bm \Sigma}_{-k} [\bar {\bm V}_{-k1}, \bar{\bm V}_{-k0}]^{\text H}$, by rearranging the order of the singular values of $\mathcal T(\bm H_{-k})$, it is easy to achieve
\begin{equation*}
  \bar{\bm V}_{-k0} = \begin{bmatrix}
  \text{Re}(\bm V_{-k0}) &  -\text{Im} (\bm V_{-k0})  \\
   \text{Im}(\bm V_{-k0}) & \text{Re} (\bm V_{-k0})
  \end{bmatrix}  = \mathcal T(\bm V_{-k0}).
\end{equation*}

According to Lemma \ref{lem:t_trans}, we have
\begin{equation}
\begin{split}
  R_{\text{WL-BD},k} &= \frac{1}{2} \log_2\det \left [ \bm I_{2M_k}  + \frac{P_{\text T}}{M\sigma^2_n} \tilde{\bm H}_k \mathcal T(\bm V_{-k0}) \mathcal T(\bm V_{-k0})^{\text T} \tilde{\bm H}_k^{\text T}  \right ] \\
   &= \frac{1}{2} \log_2\det {\color{red} \mathcal T} \left ( \bm I_{M_k}  + \frac{P_{\text T}}{M\sigma^2_n} {\bm H_k} \bm V_{-k0} \bm V_{-k0}^{\text H} {\bm H}_k^{\text H}  \right )  \\
   &= \log_2\det \left ( \bm I_{M_k}  + \frac{P_{\text T}}{M\sigma^2_n} {\bm H_k} \bm V_{-k0} \bm V_{-k0}^{\text H} {\bm H}_k^{\text H}  \right )  \\
   & = R_{\text{BD},k}.
  \end{split}
\end{equation}
Thus Proposition \ref{pro:dr_no_iqi} is proved.

\bibliographystyle{IEEEtran}
\bibliography{wl}

\end{document}